\documentclass[12pt,peerreview,onecolumn]{IEEEtranTCOM}
\pdfoutput=1
\usepackage{amsmath,epsfig,amssymb,verbatim,amsopn,subfigure,color}
\usepackage{cite,xspace}
\usepackage{array,algorithm,algorithmic}
\usepackage{bbm}
\usepackage{array}
\usepackage{multirow}
\usepackage{multicol}
\usepackage{rotating}
\usepackage{dsfont,float}
\usepackage{stfloats}
\usepackage{enumerate,makecell}
\usepackage{xspace}
\normalsize

\ifCLASSOPTIONpeerreview
\renewcommand{\baselinestretch}{1.8}
\fi

\ifCLASSOPTIONonecolumn
    
\else
    
\fi

\newcommand{\redcom}[1]{{\color{black}#1}\xspace}

\makeatletter
\let\l@ENGLISH\l@english
\makeatother

\makeatletter
\renewcommand*{\@opargbegintheorem}[3]{\trivlist
  \item[\hskip \labelsep{\itshape #1\ #2}] {\itshape (#3):} {\normalfont}}
\makeatother
\usepackage{relsize}
\newcommand{\mgf}  {\mathcal{M}_{I}(s)}
\newcommand{\totalmgf} {\mathcal{M}_{I_{\textrm{agg}}}(s)}
\newcommand{\fr}{f_{R}(r)}

\newcommand{\fg}{f_{G}(g)}
\newcommand{\fh}{f_{H}(h)}

\newcommand{\rad}{W}
\newcommand{\dist}{D}
\newcommand{\region}{\mathcal{A}'}
\graphicspath{{figs/},{Figures/}}

\newcommand{\A}{\frac{\beta}{\rho_0}}
\newcommand{\Bnew}{\frac{\beta r_0^{\alpha}}{P_{T_0}}}
\newcommand{\moment}{\mu_I(n)}
\newcommand{\meannode}{\overline{M}_{\textrm{active}}}
\newcommand{\aggintf}{I_{\textrm{agg}}}
\newcommand{\Pout}{P_{\textrm{out}}}
\newtheorem{remark}{Remark}
\newtheorem{theorem}{Theorem}
\newtheorem{corollary}{Corollary}
\graphicspath{{figs/},{Figures/}}

\newcommand{\AuthorOne}{Jing~Guo, {\em{Student Member, IEEE}}}
\newcommand{\AuthorTwo}{Salman~Durrani, {\em{Senior Member, IEEE}}}
\newcommand{\AuthorThree}{Xiangyun~Zhou, {\em{Member, IEEE}}}
\newcommand{\ThankOne}{The authors are with the Research School of Engineering, College of Engineering and Computer Science, The Australian National University, Canberra, ACT 0200, Australia.
Emails: \{jing.guo, salman.durrani, xiangyun.zhou\}@anu.edu.au.}

\begin{document}

\title{Performance Analysis of Arbitrarily-Shaped Underlay Cognitive Networks: \\Effect\redcom{s} of Secondary User Activity Protocols}
\author{\authorblockN{\AuthorOne,~\AuthorTwo~and \AuthorThree\thanks{\ThankOne}}}
\maketitle
%
%\vspace{+5mm}

%
%----------------------------------------------------------------------
\begin{abstract}
This paper analyzes the performance of the primary and secondary users (SUs) in an arbitrarily-shaped underlay cognitive network. In order to meet the interference threshold requirement for a primary receiver (PU-Rx) at an arbitrary location, we consider different SU activity protocols which limit the number of active SUs. We propose a framework, based on the moment generating function (MGF) of the interference due to a random SU, to analytically compute the outage probability in the primary network, as well as the average number of active SUs in the secondary network. We also propose a cooperation-based SU activity protocol in the underlay cognitive network which includes the existing threshold-based protocol as a special case. \redcom{We study the average number of active SUs for the different SU activity protocols, subject to a given outage probability constraint at the PU} and we employ it as an analytical approach to compare the effect of different SU activity protocols on the performance of the primary and secondary networks.
\end{abstract}

% A tradeoff analysis between the outage probability in the primary network and the average number of active SUs is provided in this paper

%
%%%----------------------------------------------------------------------
\begin{IEEEkeywords}
Underlay cognitive network, aggregate interference, outage probability, stochastic geometry, secondary user activity protocol.
\end{IEEEkeywords}

\ifCLASSOPTIONpeerreview
    \newpage
\fi

%%----------------------------------------------------------------------
\section{Introduction}
Cognitive radio networks is a promising technology to address the spectrum scarcity and the inefficient spectrum usage of present wireless systems~\cite{FCC-2002,Haykin-2005,Zhao-2007,Wang2011}. Cognitive radio networks allow unlicensed secondary users (SUs) access to the spectrum of the licensed primary users (PUs), without impairing the performance of the PUs. Depending on the spectrum access strategy, there are three main cognitive radio network paradigms: interweave, underlay and overlay~\cite{Goldsmith2009}. In the \textit{interweave cognitive networks}, the SUs are not allowed to cause any interference to the PUs. Thus, SUs must periodically sense the environment to detect spectrum occupancy and transmit opportunistically only when the PUs are silent~\cite{Ghasemi2008,Mai2008,Chen2009,Husheng2010,Derakhshani2012,Khoshkholgh2013b}. In the \textit{underlay cognitive networks}, SUs can concurrently use the spectrum occupied by a PU by guaranteeing that the interference at the PU is below some acceptable threshold. Thus, SUs must know the channel strengths to the PUs and are also allowed to communicate with each other in order to sense how much interference is being created to the PUs~\cite{Lee2012,Mahmood2012,Vijayandran2012,Rabbachin2011,Kahlon2011,Le2008,Kim2008,Le2014}. In the \textit{overlay cognitive networks}, there is tight interaction and active cooperation between the PUs and the SUs. Thus, SUs use sophisticated signal processing and coding to maintain or improve the PU transmissions while also obtaining some additional bandwidth for their own transmission~\cite{Khoshkholgh2013,Zhai2013,Chraiti2013}. Note that hybrid spectrum access strategies, appropriately combining the above three paradigms, have also been proposed~\cite{Kusaladharma2013,Jiang2013,Chu2014}.

For underlay cognitive networks, which are considered in this paper, it is very important to investigate the interference arising from the SUs to a PU. This interference impacts the outage probability at a PU, which is the probability that the signal to interference plus noise ratio (SINR) falls below a given threshold. The interference and outage in underlay cognitive networks has been recently investigated in the literature~\cite{Lee2012,Mahmood2012,Vijayandran2012,Rabbachin2011}. Specifically, the aggregate interference at a typical PU and a typical SU inside an infinite area cognitive network, taking the exclusion region around PUs into account, were presented in~\cite{Lee2012} and bounds on the outage probability with Rayleigh fading channels were also derived. The closed-form results for the moment generating function (MGF) of the aggregate interference and the mean interference at an annulus-center-located PU were derived in~\cite{Mahmood2012}. A framework for characterizing the aggregate interference in cognitive networks was proposed in~\cite{Vijayandran2012} for the disk region under the Rayleigh fading channel assumption. Therein, closed-form results were obtained for the special case that the path-loss exponent is $2$ or $4$ with an arbitrary location of PU inside a disk region. However, in practice the shape of network can be arbitrary and need not be a disk only. This is especially relevant for emerging ultra-dense small cell deployment scenarios~\cite{hossain2014}. In addition, the PU may be located anywhere inside the network region. When SUs are confined within an arbitrarily-shaped finite region, the aggregate interference and the outage probability are strongly influenced by the shape of region and the position of the PU. In this context, a method of calculating the approximation of $n$-th cumulant inside a non-circular region by dividing the areas into infinitesimal circular sections was suggested in~\cite{Rabbachin2011} but no explicit formulation was provided. Therefore, it is still largely an open research problem to find general frameworks for analyzing the interference and outage in arbitrarily-shaped finite underlay cognitive networks.

In underlay cognitive networks, there \redcom{are} several ways to control the interference generated by the SUs in order to satisfy the interference threshold, e.g., using multiple antennas to guide the SU signals away from the PU~\cite{Le2014}, using resource (i.e., rate and power) allocation among the SUs~\cite{Le2008} or using spread spectrum techniques to spread the SU signals below the noise floor~\cite{Kim2008}. Perhaps the simplest solution to control the interference generated by the SUs, which is considered in this work, is to employ the SU activity protocols, i.e., to simply limit the number of active SUs~\cite{Lee2012,Mahmood2012,Rabbachin2011}. In this context,~\cite{Lee2012,Mahmood2012} considered an exclusion or guard zone around the PUs, within which SUs are not allowed to transmit. A threshold-based protocol was proposed in~\cite{Rabbachin2011}, where the activity of each SU depends on the instantaneous power received at the SU from the PU. It must be noted that if no activity constraint is imposed on SUs then this is equivalent to the well-studied case of wireless ad hoc networks where all users can transmit~\cite{Srinivasa2007,Salbaroli-2009,Guo-2013}.

In this paper, we propose a general framework for analyzing the performance of arbitrarily-shaped underlay cognitive networks, with arbitrary location of the PU and different SU activity protocols. We make the following major contributions in this paper:
\begin{itemize}
\item \redcom{We utilize cooperation among SUs in underlay cognitive networks to come up with a cooperation-based SU activity protocol}. This protocol utilizes the local information exchange among SUs and includes the threshold-based protocol as a special case. We derive approximate yet accurate expressions for the MGF and the $n$-th cumulant of the aggregate interference from SUs with the cooperation-based protocol.

\item We derive the general expressions for the MGF and $n$-th cumulant of the aggregate interference at an arbitrarily located PU inside an arbitrarily-shaped region for the existing SU activity protocols. We show that many existing closed-form results in the literature for the interference analysis in the primary network can be obtained as special cases in our framework. In addition, we derive a closed-form result for the average number of active SUs. To the best of our knowledge, this is the first time that such a result has been obtained in the literature for the case of underlay cognitive network.

\item \redcom{We study the average number of active SUs for the different SU activity protocols, subject to a given outage probability constraint at the PU. We show that the guard zone protocol supports the highest number of active SUs, followed by the proposed cooperation-based protocol and then the threshold-based protocol. The advantage of the cooperation-based protocol over the guard zone protocol is that it relies on the SUs only knowing the instantaneous channel strengths to the PUs.}

% We propose a tradeoff analysis between the outage probability at the PU and the average number of active SUs, and employ it as a method to compare the performance of the primary and secondary networks for different SU activity protocols. Our comparison shows that the guard zone protocol performs the best, followed by the proposed cooperation-based protocol which outperforms the threshold-based protocol. The advantage of the cooperation-based protocol over the guard zone protocol is that it relies on the SUs only knowing the instantaneous channel strengths to the PUs and hence is more applicable in practice.
\end{itemize}

The remainder of the paper is organized as follows. Section~\ref{systemmodel} presents the detailed system model and assumptions and describes the three different SU activity protocols, including the proposed cooperation-based protocol. The proposed mathematical framework is presented in Section~\ref{formulation}. The analysis for the interference and the average number of active SUs is presented in Section~\ref{analysis} and Section~\ref{section_averagenode}, respectively. Numerical and simulation results to study the aggregate interference, outage probability and average number of active SUs are discussed in Section~\ref{section-result}. Finally, conclusions are presented in Section~\ref{conclusion}.

The following notation is used in the paper. The probability distribution function (PDF) and the cumulative distribution function (CDF) of a random variable (RV) $Z$ are represented by $f_Z(\cdot)$ and $F_Z(\cdot)$, respectively. $\mathbb{E}_Z\{\cdot\}$ denotes the expectation with respect to random variable $Z$ and $\mathbb{E}_Z^{pt}\left\{n,l,u\right\}=\int_{l}^{u}z^{n}f_Z(z) dz$ represents the $n$-th order partial moment of RV $Z$ calculated within the interval $[l,u]$. $\mathcal{M}_{Z}(s)$ is the moment generating function of the RV $Z$ with PDF $f_Z(\cdot)$. $\mu_Z(n)$ and $\kappa_Z(n)$ denote the $n$-th moment and $n$-th cumulant of a RV $Z$ respectively. Additionally, $\setminus $ is the set exclusion operator and \redcom{$\Pr(\cdot)$ indicates the probability measure}. \redcom{$\Gamma[x]=\int_0^{\infty}t^{x-1}\exp(-t) dt$ and $\Gamma[a,x]=\int_a^{\infty}t^{x-1}\exp(-t) dt$ are the complete gamma function and the incomplete upper gamma functions, respectively~\cite{gradshteyn2007}. $\Gamma(a,x_1,x_2)= \Gamma(a,x_1)-\Gamma(a,x_2)$ is the generalized incomplete gamma function and $_2F_1[\cdot,\cdot;\cdot;\cdot]$ is the Gaussian or ordinary hypergeometric function~\cite{gradshteyn2007}}.

\section{System Model}\label{systemmodel}
We consider an underlay cognitive network with a PU link, comprising of a PU transmitter (PU-Tx) and a PU receiver (PU-Rx) separated by a distance $r_0$, and $M$ SUs. The network region $\mathcal{A}$ is an arbitrarily-shaped finite region, where $\mathcal{A}\subset\mathbb{R}^2 $ and $\mathbb{R}^2$ denotes the two-dimensional Euclidean space. We do not place any restriction on the location of the PU-Tx and PU-Rx and they can be located anywhere inside the network region $\mathcal{A}$. A primary exclusion zone $\mathcal{B}$, with radius $\epsilon$, is formed around the PU-Rx and no active user is allowed to enter this region~\cite{Maivu-2009}. The SU locations are modeled according to a uniform Binomial Point Process, i.e., the $M$ SUs are independently and uniformly distributed (i.u.d.) at random inside the region $\region$, where $\region=\mathcal{A}\setminus \mathcal{B}$.

The SUs decide whether to transmit or not depending on the adopted SU activity protocol (discussed later in this section). We assume that all the nodes operate in the frequency division duplex mode. Similar to~\cite{Rabbachin2011}, we assume that in order to know the channel strength to PU-Rx each SU receives a signal transmitted by PU-Rx via a sensing channel. We assume that this sensing channel (from PU-Rx to SU) and the SU transmitting (i.e., interfering) channel (from SU to PU-Rx) are well separated in the frequency band so that these two channels can be regarded as fully uncorrelated.
\ifCLASSOPTIONonecolumn
\begin{figure}
\centering
\subfigure[Guard zone protocol with radius $r_f$ and primary exclusion zone radius $\epsilon$.]{\label{guardzone}\includegraphics[width=0.44\textwidth]{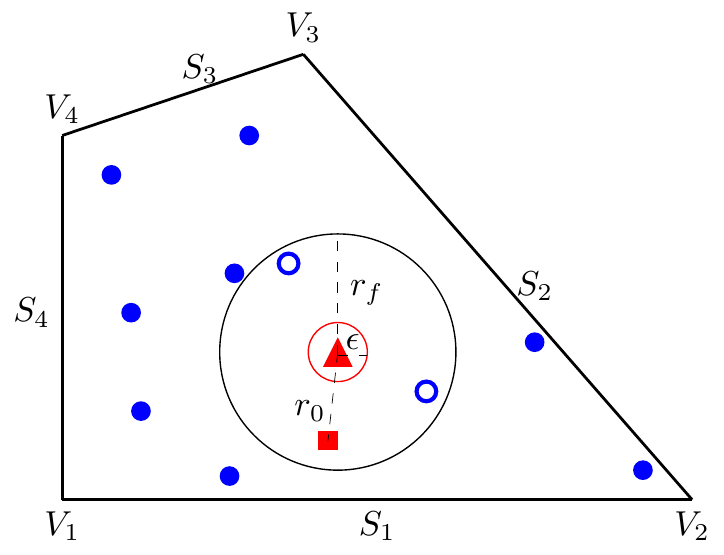}}
\\
\subfigure[Threshold-based protocol with activation threshold $\gamma$.]{\label{singlethreshold}\includegraphics[width=0.44\textwidth]{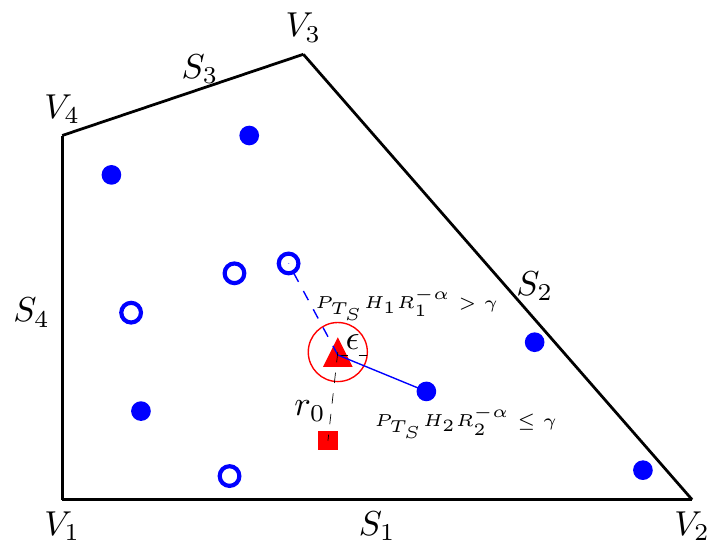}}
%\mbox{\hspace{0.5cm}}
\subfigure[Cooperation-based protocol with activation threshold $\gamma$ and cooperation range $r_c$ .]{ \label{coop}\includegraphics[width=0.44\textwidth]{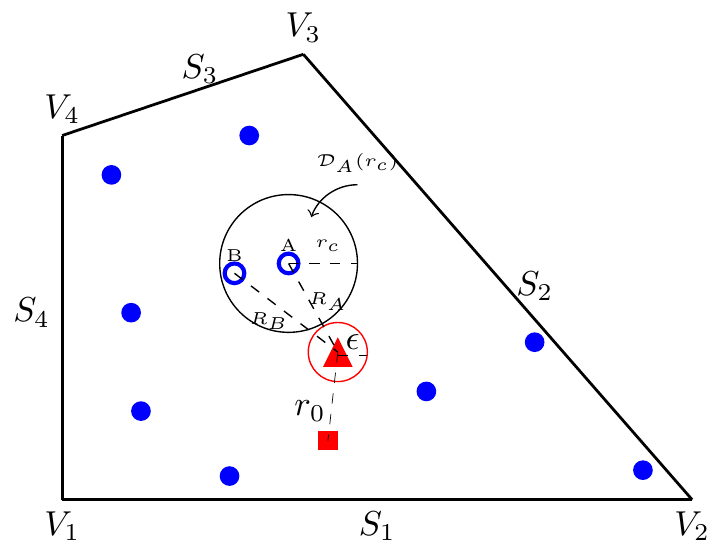}}
\caption{Illustration of secondary user spatial activity protocols in underlay cognitive network (${\color{red}\blacktriangle}=$ interfered PU-Rx, ${\color{red}\blacksquare}=$ PU-Tx, ${\color{blue}\circ}=$ inactive secondary user, ${\color{blue}\bullet}=$ active secondary user). $S_j$ and $V_j$ ($j=1,2,3,4$) denote the side and vertex, respectively.}
\end{figure}
\else
\begin{figure}
\center
\subfigure[Guard zone protocol with radius $r_f$ and primary exclusion zone radius $\epsilon$.]{\label{guardzone}\includegraphics[width=0.4\textwidth]{fig/fixedsensing}}
\\
\subfigure[Threshold-based protocol with activation threshold $\gamma$.]{\label{singlethreshold}\includegraphics[width=0.4\textwidth]{fig/dynamicsensing}}
\\%\mbox{\hspace{0.5cm}}
\subfigure[Cooperation-based protocol with activation threshold $\gamma$ and cooperation range $r_c$ .]{ \label{coop}\includegraphics[width=0.4\textwidth]{fig/coop}}
\caption{Illustration of secondary user spatial activity protocols in underlay cognitive network (${\color{red}\blacktriangle}=$ interfered PU-Rx, ${\color{red}\blacksquare}=$ PU-Tx, ${\color{blue}\circ}=$ inactive secondary user, ${\color{blue}\bullet}=$ active secondary user). $S_j$ and $V_j$ ($j=1,2,3,4$) denote the side and vertex, respectively.}
\end{figure}
\fi

Let the RV $R_i$ $(i=1, 2,\ldots, M)$ denote the random distance between the $i$-th SU and the PU-Rx with probability distribution function $f_{R_i}(r_i)$. We denote the transmit power of the PU-Tx as $P_{T_0}$, the transmit power of each SU as $P_{T_i}$ and the transmit power of PU-Rx as $P_{T_S}$. We assume that all users have a single antenna and the wireless communication channel is modeled as a path-loss and fading channel. Let $G_i$ represent the instantaneous power gain due to fading on the SU transmitting channel from $i$-th SU to the PU-Rx with fading distribution function $f_{G_i}(g_i)$ and $H_i$ represent the instantaneous fading power gain on the sensing channel with the distribution function $f_{H_i}(h_i)$.

For the above setup, the interference at the PU-Rx generated from the $i$-th SU is given by
 \begin{align}\label{inf}
I_i=P_{T_i}G_i R_i^{-\alpha}\textbf{1}_{(\textrm{condition})},
\end{align}
\noindent where $\alpha$ is the path-loss exponent which is typically in the range $2\leq\alpha\leq6$~\cite{Marvin2005}. The indicator function is given by
\begin{align}
\textbf{1}_{(\textrm{condition})}=\begin{cases}
              1, & {\textrm{ if condition is true};} \\
            0, & {\textrm{else if condition is false};}
             \end{cases}
\end{align}

\noindent where the ``condition" depends on the different SU activity protocols and the explicit expressions for each protocol are given in~\eqref{intf_guard},~\eqref{single} and~\eqref{intf_coop}, respectively. In addition, although the unbounded path-loss model is used in~\eqref{inf}, the singularity at $R_i=0$ and the amplification of the transmitted signal are avoided because of the primary exclusion zone around the PU-Rx, i.e., the random distance $R_i$ is always greater than $\epsilon$ ($\epsilon\geq1$)~\cite{Ghasemi2008}.

The aggregate interference at PU-Rx is given by
 \begin{align}\label{total_inf}
\aggintf=\sum_{i=1}^M I_i=\sum_{i=1}^{M}P_{T_i}G_i R_i^{-\alpha}\textbf{1}_{(\textrm{condition})}.
\end{align}

In the following subsections, we present the definition of each SU activity protocol.

%\subsection{Full Activity Protocol}\label{sec:fullactivity}
%This is an unregulated activity protocol, in which all the SUs are in active status and share the same spectrum occupied by the PU-Rx. This is depicted in Fig.~\ref{network}. Since all the SUs are active, the indicator function in~\eqref{inf} and~\eqref{total_inf} is always equal to one and we can rewrite the aggregate interference in~\eqref{total_inf} as
%%
%\begin{align}\label{inf_full}
%\aggintf=\sum_{i=1}^{M}P_{T_i}G_i R_i^{-\alpha}.
%\end{align}

\subsection{Guard Zone Protocol}\label{sec:guardzone}
The guard zone protocol was employed in~\cite{Lee2012,Mahmood2012}. In this protocol, the SUs are permitted to enter the guard zone region but once a SU intrudes into it, it is prohibited from transmitting. This is illustrated in Fig.~\ref{guardzone}, where there is a guard zone region around the PU-Rx with radius $r_f$. Consequently, the two SUs that are inside this region are inactive and do not generate any interference to the PU-Rx. The aggregate interference under the guard zone protocol can be written as
\begin{align}\label{intf_guard}
\aggintf=\sum_{i=1}^{M}P_{T_i}G_i R_i^{-\alpha}\textbf{1}_{\left(R_i>r_f\right)}.
\end{align}

\subsection{Threshold-based Protocol}\label{sec:single}
The threshold-based protocol was proposed in~\cite{Rabbachin2011}. In this protocol, each SU receives the instantaneous signal power transmitted by the PU-Rx on the sensing channel. If the received instantaneous signal power at the $i$-th SU is greater than the activation threshold $\gamma$, i.e., $P_{T_S}H_iR_i^{-\alpha}>\gamma$, it becomes silent and does not interfere with the PU-Rx. Otherwise, it is permitted to transmit, as illustrated in Fig.~\ref{singlethreshold}. Hence, for the threshold-based protocol, the aggregate interference can be written as
\begin{align}\label{single}
\aggintf=\sum_{i=1}^{M}P_{T_i}G_i R_i^{-\alpha}\textbf{1}_{\left(P_{T_S}H_i R_i^{-\alpha}\leq\gamma\right)}.
\end{align}

\subsection{Cooperation-based Protocol}\label{sec:cooperative}
This is the new protocol proposed in this paper and is illustrated in Fig.~\ref{coop}. The basic idea of this protocol is inspired from the cooperative spectrum sensing in interweave cognitive networks, where cooperation among nodes helps to improve the detection of licensed spectrum occupancy~\cite{Ghasemi2008,Derakhshani2012}.\footnote{The notion of cooperation among SUs is also similar in spirit to base station cooperation in cellular networks~\cite{Akoum2013}.}

In the proposed cooperation-based protocol for underlay cognitive networks, each SU receives the instantaneous signal power transmitted by the PU-Rx on the sensing channel and forms an initial decision on activation. Then this initial decision is broadcast to other SUs. For analytical convenience, we assume that, for each SU, it can only correctly receive the initial decisions from other SUs within a certain range, which is known as its cooperation range $r_{c}$. Finally, in order to decide whether it is active or not, each SU applies the AND rule on the received initial decisions from other cooperating SUs and its own initial decision. Consequently, for a considered SU, it is permitted to be active as long as its preliminary decision is to be active, and the initial decision of all SUs which fall into this SU's cooperation range is also to be active. Mathematically, the aggregate interference generated at the PU-Rx is
\begin{align}\label{intf_coop}
\aggintf=\sum_{i=1}^{M}P_{T_i}G_i R_i^{-\alpha}\textbf{1}_{\left(\Pi_{d}\left(\mathcal{D}_{i}\left(r_c\right)\times\region\right)=\emptyset\right)},
\end{align}
\noindent where $\Pi_{d}$ denotes the set of SUs whose received instantaneous signal power on the sensing channel is greater than the activation threshold $\gamma$, $\mathcal{D}_{i}\left(r_c\right)$ represents the disk cooperation region centered at the $i$-th SU and $\emptyset$ denotes the null set. Note that when $r_c=0$, the cooperation-based protocol is the same as the threshold-based protocol. Thus, the proposed cooperation-based protocol includes the threshold-based protocol as a special case.

\begin{remark}
Both the cooperation-based and threshold-based protocols require the SU to receive the instantaneous signal power transmitted by the PU-Rx on the sensing channel. As such, they are much more applicable in practice. However, the guard zone protocol requires the SU to know the instantaneous signal power on the sensing channel over a relatively long period of time and then average it to determine its distance to the PU-Rx, before deciding whether to transmit or not.\footnote{Alternatively, the guard zone protocol can also be implemented using cooperative localization techniques~\cite{Wymeersch2009}.} As such, this protocol is not suitable for the scenarios where the SUs need to transmit without too much delay.
\end{remark}

%%***************************************************************************%%
\section{Mathematical Framework}\label{formulation}
In this section, we present the proposed mathematical framework to characterize the interference and outage in underlay cognitive networks with different SU activity protocols. The aggregate interference from the secondary network in~\eqref{total_inf} is a stochastic process that strongly relies on the random location of the SUs inside the arbitrarily-shaped finite cognitive network region and the random fading channel gains. Since there is no available general expression for the PDF of the aggregate interference~\cite{Haenggi-2012,ElSawy-2013}, we adopt the moment generating function approach to analyze the interference and outage in this paper. Previous work~\cite{Rabbachin2011,Vijayandran2012} has also adopted the MGF approach. However, their focus is on analysing the statistics of the aggregate interference in the primary network only and the results are limited to specific, e.g., annulus-shaped regions. We consider arbitrarily-shaped cognitive network regions and analyze the performance in both the primary network (i.e., the aggregate interference in Section~\ref{analysis}) and the secondary network (i.e., the average number of active SUs in Section~\ref{section_averagenode}).

\subsection{Assumptions}
In this paper, we consider that the nodes are independently and uniformly distributed inside the network region $\region$, which results in the distribution function of $R_i$ being the same for all $i$. Moreover, the fading gain on all communication channels is independently and identically distributed (i.i.d.) Nakagami-$m$ fading. This type of fading is widely considered in the wireless communications literature~\cite{Marvin2005}. The transmit power for different SUs are assumed to be the same. Consequently, the distribution of interference from $i$-th SU becomes identical and we can drop the index $i$ in the $P_{T_i}$, $I_i$, $R_i$, $G_i$, $H_i$ and let $f_{R_i}(r_i)=f_{R}(r)$, $f_{G_i}(g_i)=f_{G}(g)$ and $f_{H_i}(h_i)=f_{H}(h)$.

For Nakagami-$m$ fading, the distribution of the power gain on the SU transmitting channel and the sensing channel can be modeled by a Gamma distribution as~\cite{Marvin2005}
\begin{align}\label{gamma_distr}
f_{G}(g)=&\frac{g^{m_g-1}m_g^{m_g}}{\Gamma[m_g]}\exp(-m_g g),\\
f_{H}(h)=&\frac{h^{m_h-1}m_h^{m_h}}{\Gamma[m_h]}\exp(-m_h h),
\end{align}
\noindent where $m_g$ and $m_h$ represent the fading parameters, which control the severity of the fading. Note that $m_g=m_h=1$ corresponds to Rayleigh fading channels. In addition, the $n$-th moment of the fading power gain on the SU transmitting channel, which is needed in the analysis in Section~\ref{analysis}, is available in closed-form as~\cite{Nakagami1960}
\begin{align}\label{gamma_expectation}
\mathbb{E}_G\left\{G^n\right\}=&\frac{(m_g+n-1)!}{m_g^n(m_g-1)!}.
\end{align}

\subsection{Distance Distributions}
The proposed formulation relies on the knowledge of the distance distribution $\fr$, i.e., the PDF of the distance of a random SU from the PU-Rx. For a disk region, $\fr$ is well known~\cite{Rabbachin2011,Vijayandran2012,Mahmood2012,Mai2008,Derakhshani2012,Srinivasa2007}. For an $L$-sided arbitrarily-shaped convex polygon, $\fr$ can be a complicated piece-wise function with at most $2L$ piece-wise terms~\cite{Zubair-2013}. The number of piece-wise terms depends on the number of unique distances between the location of the reference node and all the sides and vertices, respectively, of the polygon region. Recently,~\cite{Guo-2013} proposed an algorithm to determine $\fr$ for the case of a random node located anywhere inside an arbitrarily-shaped convex polygon. This algorithm is used in this work to determine $\fr$ in closed-form. Once $\fr$ is determined using the algorithm in~\cite{Guo-2013}, the expectation $\mathbb{E}_R\left\{R^{-n\alpha}\right\}$ involving the RV $R$, which is needed in the analysis in Section~\ref{analysis}, can be easily calculated in closed-form.

\subsection{Moment Generating Function}
In general, the moment generating function of the aggregate interference is defined as~\cite{Marvin2005}
\begin{align}\label{mgf}
\totalmgf = \mathbb{E}_{\aggintf}\left\{\exp\left(-s \aggintf\right)\right\},
\end{align}
\noindent where $\mathbb{E}_{\aggintf}\{\cdot\}$ denotes the expectation with respect to the RV $\aggintf$. Assuming that the interference from each SU is independent and identical, the moment generating function of the aggregate interference in~\eqref{mgf} can be rewritten as~\cite{Marvin2005}
\begin{align}\label{mgf-bpp}
\totalmgf=\left(\mgf \right)^M,
\end{align}
\noindent where $I$ denotes the interference generated by a random SU and $\mgf=\mathbb{E}_{I}\left\{\exp(-s I)\right\}$ corresponds to the MGF of $I$.

\subsection{$n$-th Cumulant}\label{section:cumulant}
The $n$-th cumulant of the aggregate interference can be written in terms of the MGF of the aggregate interference as~\cite{Marvin2005}
\begin{align}\label{cu-bpp}
\kappa_{\aggintf}(n)=& (-1)^n\left.\frac{d^n \ln\totalmgf}{d s^n}\right|_{s=0}\nonumber\\
            =&(-1)^nM\left.\frac{d^n \ln\mgf}{d s^n}\right|_{s=0}\nonumber\\
                                     =& M \kappa_{I}(n) \nonumber\\
                                     =&M\left(\mu_I(n)-\sum_{j=1}^{n-1}\binom{n-1}{j-1}\kappa_I(j)\mu_I(n-j)\right),
\end{align}
\noindent where the last step comes from the recursive moment-cumulant relationship~\cite{Srinivasa2007}, and $\kappa_I(n)$ and $\mu_I(n)$ represent the $n$-th cumulant and $n$-th moment of the interference from a random SU respectively. Note that $\mu_I(n)$ can also be directly related to $\mgf$ by~\cite{Marvin2005}
\begin{align}\label{moment_relation}
\moment=(-1)^n\left.\frac{d^n \mgf}{d s^n}\right|_{s=0}.
\end{align}

\subsection{Outage Probability}\label{section:outage_calculation}
The outage probability is an important metric to evaluate the impact of SU activity protocols on the performance of the primary users over fading channels. It is given by
\begin{align}\label{outage}
P_{\textrm{out}}=\Pr\left(\textrm{SINR}<\beta\right)=\Pr\left(\frac{P_{T_0}r_0^{-\alpha}G_0}{N+\aggintf}<\beta\right),
\end{align}
\noindent where $\beta$ is the SINR threshold and $N$ is the additive white Gaussian noise power.

In this paper, we are interested in the spatially averaged outage probability which is spatially averaged over both the possible location of the SUs and the fading channels. When the fading on the desired link (from PU-Tx to PU-Rx) follows the general distribution defined in~\cite[(9)]{Hunter-2008} (an important special case of which is Nakagami-$m$ fading with integer $m$ value), we can employ the reference link power gain-based (RLPG-based) framework proposed in~\cite{Guo-2013} to evaluate the spatially averaged outage probability. The basic principle of this approach is to first condition on the interference and express the outage probability in terms of the CDF of the reference link's fading power gain. The conditioning on the interference is then removed first by removing the conditioning on the fading power gains of the interferers and then removing the conditioning on the locations of the interferers. For Nakagami-$m$ fading channels, the spatially averaged outage probability is given by~\cite{Guo-2013}
\ifCLASSOPTIONonecolumn
\begin{small}
\begin{align}\label{outage_bpp}
\Pout =1-\exp\left(-m_0\A\right)\sum_{k=0}^{m_0-1}\frac{m_0^k}{k!}\sum_{j=0}^{k}\binom{k}{j}\left(\A\right)^{k-j}\left(\Bnew\right)^j \sum_{t_1+t_2...+t_M=j}\binom{j}{t_1,t_2,...,t_M}\prod_{i=1}^{M}\mathbb{E}_{I}\left\{\exp\left(-m_0\Bnew I\right)\left(I\right)^{t_i}\right\},
\end{align}
\end{small}
\else

\begin{small}
\begin{align}\label{outage_bpp}
\Pout =&1-\exp\left(-m_0\A\right)\sum_{k=0}^{m_0-1}\frac{m_0^k}{k!}\sum_{j=0}^{k}\binom{k}{j}\left(\A\right)^{k-j}\left(\Bnew\right)^j \nonumber\\
&\sum_{t_1+t_2...+t_M=j}\binom{j}{t_1,t_2,...,t_M}\prod_{i=1}^{M}\mathbb{E}_{I}\left\{\exp\left(-m_0\Bnew I\right)\left(I\right)^{t_i}\right\},
\end{align}
\end{small}
\fi

\noindent where $\rho_0=\frac{P_{T_0}r_0^{-\alpha}}{N}$ indicates the average signal-to-noise ratio (SNR) and $m_0$ denotes the fading parameter on the desired link.

Using the frequency differentiation property of Laplace transform, the Laplace transform of the term $z^n f(z)$ is given by $(-1)^n\frac{d^{n} \mathcal{M}_{Z}(s)}{d s^{n}}$. Then the expectation term in~\eqref{outage_bpp} can be expressed in terms of the MGF of the interference due to a random SU as
\begin{align}\label{outage_bpp1}
\mathbb{E}_{I}\left\{\exp\left(-m_0\Bnew I\right)\left(I\right)^{t_i}\right\}=(-1)^{t_i}\left.\frac{d^{t_i} \mgf}{d s^{t_i}}\right|_{s=m_0\Bnew}.
\end{align}

% For the case that the desired link is impaired by other types of fading, the inversion of moment generating function can be adopted. The detail numerical inversion procedure is available in~\cite{Ko-2000},~\cite{Guo-2013} and we are not going to cover it in this paper.

Examining~\eqref{mgf-bpp},~\eqref{cu-bpp} and~\eqref{outage_bpp}, we can see that the proposed mathematical formulation depends on the MGF of the interference due to a random SU $\mgf$. This is determined for the different SU activity protocols in the next section.

\section{Interference Analysis}\label{analysis}
In this section, we derive the general expressions characterizing the MGF of the interference from a random SU for the different SU activity protocols.

\subsection{Guard Zone Protocol}
For this protocol, the SUs within the guard zone region do not transmit. The main result is summarized in Theorem~\ref{theorem_guard}.
\begin{theorem}\label{theorem_guard}
For the guard zone protocol, the MGF of the interference at an arbitrarily located PU-Rx due to an independently and uniformly distributed SU inside an arbitrarily-shaped finite region is
\ifCLASSOPTIONonecolumn
\begin{align}\label{mgf_fix}
\mgf=\int_0^{\infty}\int_{r_f}^{r_{\textrm{max}}}\exp\left(-s P_{T}g r^{-\alpha}\right)\fr \fg dr dg +F_R(r_f),
\end{align}
\else
\begin{align}\label{mgf_fix}
\mgf=&\int_0^{\infty}\int_{r_f}^{r_{\textrm{max}}}\exp\left(-s P_{T}g r^{-\alpha}\right)\fr \fg dr dg \nonumber\\
&+F_R(r_f),
\end{align}
\fi

\noindent where $F_R(\cdot)$ represents the cumulative distribution function of the distance of a random SU from the PU-Rx, which can be determined by the algorithm in~\cite{Guo-2013}.
\end{theorem}

\begin{corollary}\label{corollary_guard}
For the guard zone protocol, the $n$-th moment of the interference at an arbitrarily located PU-Rx due to an independently and uniformly distributed SU inside an arbitrarily-shaped finite region is
\begin{align}\label{moment_fix}
\moment=P_{T}^n \mathbb{E}_G\left\{G^n\right\}\mathbb{E}_R^{pt}\left\{-n\alpha,r_f,r_{\textrm{max}}\right\}.
\end{align}
\end{corollary}

Proof of Theorem~\ref{theorem_guard} and Corollary~\ref{theorem_guard}: See Appendix~\ref{app_guard}.
\begin{remark}
\eqref{mgf_fix} and~\eqref{moment_fix} can also be used to obtain the results for the full activity protocol in which no activity constraint is
imposed on SUs and all the SUs are in active status. In the full activity protocol, a SU located within the maximum range of $\epsilon$ and $r_{\textrm{max}}$ generates interference to the PU-Rx. In the guard zone protocol, a SU located within the smaller range of $r_f$ and $r_{\textrm{max}}$ generates interference to the PU-Rx. Thus, when the guard zone range $r_f$ is set to equal to $\epsilon$, the guard zone protocol reduces to the full activity protocol. Therefore, the MGF and $n$-th cumulant results for the full activity protocol are the same as~\eqref{mgf_fix} and~\eqref{moment_fix} with $r_f$ replaced by $\epsilon$.
\end{remark}

\textit{Special Case of a Regular $L$-sided Polygon:} Consider the special case that the PU-Rx is located at the center of a regular $L$-sided polygon which is inscribed in a circle of radius $\rad$. In this case, the distance distribution function is given by~\cite{Zubair-2013}
\begin{align}\label{distance_distr}
\fr=\frac{1}{|\region|}  \left\{ \begin{array}{ll}
        2\pi r, &{\epsilon\leq r\leq W_p;}\\
      2\pi r -2Lr \arccos\left(\frac{W_p}{r} \right),                 &{W_p\leq r\leq\rad;}
                    \end{array} \right.
\end{align}

\noindent where $|\region|=\frac{1}{2}L\rad^2 \sin\left(\frac{2\pi}{L}\right)-\pi\epsilon^2$ denotes the area of the underlay secondary network region, $\theta =\frac{\pi(L-2)}{L}$ is the interior angle between two adjacent sides of the polygon and $W_p=\rad\sin{\left(\frac{\theta}{2}\right)}$ is the perpendicular distance from the center of the polygon to any side. Substituting~\eqref{distance_distr} and~\eqref{gamma_distr} into~\eqref{mgf_fix} and~\eqref{moment_fix}, yields the following results

\ifCLASSOPTIONonecolumn
\begin{align}\label{mgf_center_disk_fix}
\mgf=&\frac{\pi \left(\rad^2\left._2F_1\right.\left[m_g,-\frac{2}{\alpha};\frac{-2+\alpha}{\alpha};-\frac{\rad^{-\alpha}s P_T}{m_g}\right]-r_f^2\left._2F_1\right.\left[m_g,-\frac{2}{\alpha};\frac{-2+\alpha}{\alpha};-\frac{r_f^{-\alpha}s P_T}{m_g}\right]+r_f^2-\epsilon^2\right)}{|\region|}\nonumber\\
&-\int_{W_p}^{\rad}\frac{2m^m L r}{\left(m+r^{-\alpha}s P_T\right)^m}\arccos\left(\frac{W_p}{r} \right)dr,
\end{align}
\else
\begin{align}\label{mgf_center_disk_fix}
\mgf=&\frac{\pi \left( \rad^2\left._2F_1\right.\left[m_g,-\frac{2}{\alpha};\frac{-2+\alpha}{\alpha};-\frac{\rad^{-\alpha}s P_T}{m_g}\right]+r_f^2\right)}{|\region|}\nonumber\\
&-\frac{\pi \left( r_f^2\left._2F_1\right.\left[m_g,-\frac{2}{\alpha};\frac{-2+\alpha}{\alpha};-\frac{r_f^{-\alpha}s P_T}{m_g}\right]+\epsilon^2\right)}{|\region|}\nonumber\\
&-\int_{W_p}^{\rad}\frac{2m^m L r}{\left(m+r^{-\alpha}s P_T\right)^m}\arccos\left(\frac{W_p}{r} \right)dr,
\end{align}
\fi
\ifCLASSOPTIONonecolumn
\begin{align}\label{moment_center_disk_fix}
\moment=P_{T}^n \frac{(m_g+n-1)!}{m_g^n (m_g-1)!}\frac{2\left(\pi\left(\rad^{2-n\alpha}-r_f^{2-n\alpha}\right)-L\Phi\left(W\right)+L\Phi\left(W_p\right) \right)}{|\region|(2-n\alpha)},
\end{align}
\noindent where $\Phi\left(r\right)=\frac{(1-n\alpha)\left((1+n\alpha)r^3\arccos{\left(\frac{W_p}{r}\right)}+W_p^3 \left._2F_1\right.\left[\frac{1}{2},\frac{n\alpha+1}{2};\frac{n\alpha+3}{2};\frac{W_p^2}{r^2}\right] \right)-W_p(1+n\alpha)r^2\left._2F_1\right.\left[-\frac{1}{2},\frac{n\alpha-1}{2};\frac{n\alpha+1}{2};\frac{W_p^2}{r^2}\right]}{(n\alpha -1) (n\alpha+ 1)r^{1+n\alpha}}$ and $r_f$ is assumed to be less than $W_p$.
\else
\begin{align}\label{moment_center_disk_fix}
\moment=&P_{T}^n \frac{(m_g+n-1)!}{m_g^n (m_g-1)!}\times\nonumber\\
&\frac{2\left(\pi\left(\rad^{2-n\alpha}-r_f^{2-n\alpha}\right)-L\Phi\left(W\right)+L\Phi\left(W_p\right) \right)}{|\region|(2-n\alpha)},
\end{align}
\noindent where $\Phi\left(r\right)=\frac{-W_p(1+n\alpha)r^2\left._2F_1\right.\left[-\frac{1}{2},\frac{n\alpha-1}{2};\frac{n\alpha+1}{2};\frac{W_p^2}{r^2}\right] }{(n\alpha -1) (n\alpha+ 1)r^{1+n\alpha}}$ $+\frac{(1-n\alpha)\left((1+n\alpha)r^3\arccos{\left(\frac{W_p}{r}\right)}+W_p^3 \left._2F_1\right.\left[\frac{1}{2},\frac{n\alpha+1}{2};\frac{n\alpha+3}{2};\frac{W_p^2}{r^2}\right] \right)}{(n\alpha -1) (n\alpha+ 1)r^{1+n\alpha}}$ and $r_f$ is assumed to be less than $W_p$.
\fi

Note that while~\eqref{mgf_center_disk_fix} does not have a closed-form result due to the integration involving the $\arccos(\cdot)$ term, it can be easily computed numerically.

%\textit{Special Case of a Disk Region:}
\begin{remark}
When $L\rightarrow \infty$, $W_p \rightarrow W$ and the regular $L$-sided polygon approaches a disk region. The disk region or annulus-shaped region with centered PU-Rx and full activity protocol is the most popular scenario and has been widely analyzed in previous works~\cite{Srinivasa2007, Rabbachin2011, Vijayandran2012, Mahmood2012}. Under the Nakagami-$m$ fading assumption, the MGF calculated from~\eqref{mgf_center_disk_fix} by setting $r_f=\epsilon$ (i.e., full activity protocol) and $W_p=W$ (i.e., the integration term in~\eqref{mgf_center_disk_fix} reduces to zero) is identical to the result in~\cite[eq. (6)]{Srinivasa2007}. In addition, the $n$-th cumulant calculated from~\eqref{moment_center_disk_fix} (replacing $r_f$ by $\epsilon$ and $\Phi\left(W_p\right)=\Phi\left(W\right)$) and~\eqref{cu-bpp} is the same as the result from~\cite{Srinivasa2007}.
\end{remark}

\subsection{Threshold-based Protocol}
In this protocol, the activity of each SU depends on the instantaneous signal power received on the sensing channel. The main result is summarized in Theorem~\ref{theorem_single}.
\begin{theorem}\label{theorem_single}
For the threshold-based protocol, assuming the sensing channel is fully uncorrelated with the SU transmitting channel, the MGF of the interference at an arbitrarily located PU-Rx due to an independently and uniformly distributed SU inside an arbitrarily-shaped finite region is
\ifCLASSOPTIONonecolumn
 \begin{align}\label{mgf_single}
 \mgf =\int_0^{\infty}\int_{\epsilon}^{r_{\textrm{max}}}\exp\left(-s P_T g r^{-\alpha}\right)F_H\left(\frac{\gamma r^\alpha}{P_{T_S}}\right)\fr\fg dr dg+1-\int_{\epsilon}^{r_{\textrm{max}}}F_H\left(\frac{\gamma r^\alpha}{P_{T_S}}\right)\fr dr,
 \end{align}
\else
 \begin{align}\label{mgf_single}
 \mgf =&\int_0^{\infty}\int_{\epsilon}^{r_{\textrm{max}}}\exp\left(-s P_T g r^{-\alpha}\right)F_H\left(\frac{\gamma r^\alpha}{P_{T_S}}\right)\fr\fg dr dg\nonumber\\
 &+1-\int_{\epsilon}^{r_{\textrm{max}}}F_H\left(\frac{\gamma r^\alpha}{P_{T_S}}\right)\fr dr,
 \end{align}
\fi
\noindent where $F_H(\cdot)$ denotes the CDF of the fading power gain on sensing channel.
\end{theorem}

\begin{corollary}\label{corollary_single}
For the threshold-based protocol, the $n$-th moment of the interference at an arbitrarily located PU-Rx due to an independently and uniformly distributed SU inside an arbitrarily-shaped finite region is
\begin{align}\label{moment_single}
\moment=P_{T}^n \mathbb{E}_G\left\{G^n\right\}\mathbb{E}_R\left\{F_H\left(\frac{\gamma R^\alpha}{P_{T_S}}\right)R^{-n\alpha}\right\}.
\end{align}
\end{corollary}

Proof of Theorem~\ref{theorem_single} and Corollary~\ref{theorem_single}: See Appendix~\ref{app_single}.

\ifCLASSOPTIONonecolumn
\textit{Special Case of regular $L$-sided Polygon:} Substituting the PDFs $\fr$ in~\eqref{distance_distr} and $\fg$ in~\eqref{gamma_distr} into~\eqref{mgf_single} and~\eqref{moment_single}, we can obtain the MGF and the $n$-th moment for this special case. For simplicity, we only show the $n$-th moment result, which is given by
\begin{align}\label{moment_center_disk_single}
\moment=&P_{T}^n\frac{(m_g+n-1)!}{m_g^n (m_g-1)!}\frac{2\pi}{|\region|(2-n\alpha)}\left(\rad^{2-n\alpha}\frac{\Gamma[m_h,0,m_h \gamma\rad^\alpha/P_{T_S}]}{\Gamma[m_h]}\right.\nonumber\\
&\left.-\epsilon^{2-n\alpha}\frac{\Gamma[m_h,0,m_h \gamma\epsilon^\alpha/P_{T_S}]}{\Gamma[m_h]}-\left(\frac{m_h\gamma}{P_{T_S}}\right)^{n-\frac{2}{\alpha}}\frac{\Gamma[m_h-n+\frac{2}{\alpha},m_h \gamma\epsilon^\alpha/P_{T_S},m_h \gamma\rad^\alpha/P_{T_S}]}{\Gamma[m_h]}\right.\nonumber\\
&\left.-\frac{L(2-n\alpha)}{\pi}\int_{W_p}^{W}\frac{\Gamma\left[m_h,0,\frac{m_h\gamma r^\alpha}{P_{T_S}} \right]}{\Gamma[m_h]}r^{1-n\alpha} \arccos\left(\frac{W_p}{r} \right)dr\right),
\end{align}
\noindent \redcom{where $\Gamma[\cdot,\cdot,\cdot]$ is the generalized incomplete gamma function~\cite{gradshteyn2007}.}
\else
\textit{Special Case of regular $L$-sided Polygon:} Substituting the PDFs $\fr$ in~\eqref{distance_distr} and $\fg$ in~\eqref{gamma_distr} into~\eqref{mgf_single} and~\eqref{moment_single}, we can obtain the MGF and the $n$-th moment for this special case. For simplicity, we only show the $n$-th moment result, which is given by~\eqref{moment_center_disk_single} shown at the top of next page, \redcom{where $\Gamma[\cdot,\cdot,\cdot]$ is the generalized incomplete gamma function~\cite{gradshteyn2007}.}
\begin{figure*}[!t]
\normalsize
\begin{align}\label{moment_center_disk_single}
\moment=&P_{T}^n\frac{(m_g+n-1)!}{m_g^n (m_g-1)!}\frac{2\pi}{|\region|(2-n\alpha)}\left(\rad^{2-n\alpha}\frac{\Gamma[m_h,0,m_h \gamma\rad^\alpha/P_{T_S}]}{\Gamma[m_h]}\right.\nonumber\\
&\left.-\epsilon^{2-n\alpha}\frac{\Gamma[m_h,0,m_h \gamma\epsilon^\alpha/P_{T_S}]}{\Gamma[m_h]}-\left(\frac{m_h\gamma}{P_{T_S}}\right)^{n-\frac{2}{\alpha}}\frac{\Gamma[m_h-n+\frac{2}{\alpha},m_h \gamma\epsilon^\alpha/P_{T_S},m_h \gamma\rad^\alpha/P_{T_S}]}{\Gamma[m_h]}\right.\nonumber\\
&\left.-\frac{L(2-n\alpha)}{\pi}\int_{W_p}^{W}\frac{\Gamma\left[m_h,0,\frac{m_h\gamma r^\alpha}{P_{T_S}} \right]}{\Gamma[m_h]}r^{1-n\alpha} \arccos\left(\frac{W_p}{r} \right)dr\right).
\end{align}
\hrulefill
\vspace*{4pt}
\end{figure*}
\fi

\begin{remark}
For the disk region,~\eqref{moment_center_disk_single} with $L=\infty$ (i.e., ignoring the integration part in~\eqref{moment_center_disk_single}) is equivalent to the result in~\cite{Rabbachin2011}. However, the method of calculating the $n$-th cumulant in~\cite{Rabbachin2011} is only applicable for the special case that PU-Rx is located at the center of the disk region.
\end{remark}

\subsection{Cooperation-based Protocol}\label{sec:coop:analysis}
For the cooperation-based protocol, the activity of each SU is determined by itself as well as other SUs within its cooperative range. Thus, the interference due to each SU is not independent and~\eqref{mgf-bpp} is not strictly valid. The analysis in the presence of correlated interference is an important open research problem. In this paper, we still use~\eqref{mgf-bpp} to derive approximate analytical results for the cooperation-based protocol. We show that these results are accurate under certain conditions,  which will be discussed in detail in Section~\ref{section-result}. The main result is summarized in Theorem~\ref{theorem_coop} below.
\begin{theorem}\label{theorem_coop}
\ifCLASSOPTIONonecolumn
For the cooperation-based protocol, the MGF of the interference at an arbitrarily located PU-Rx due to an independently and uniformly distributed SU inside an arbitrarily-shaped finite region is approximated by
 \begin{align}\label{mgf_coop}
 \mgf \approx &\int_0^{\infty}\int_{\epsilon}^{r_{\textrm{max}}}\exp\left(-s P_{T} g r^{-\alpha}\right)F_H\left(\frac{\gamma r^\alpha}{P_{T_S}}\right)\left(\frac{\left|\region-\pi r_c^2\right|}{\left|\region\right|}
 +\frac{\pi r_c^2}{\left|\region\right|}F_H\left(\frac{\gamma r^\alpha}{P_{T_S}}\right)\right)^{M-1}\fr\fg dr dg  \nonumber\\
 &+1-\int_{\epsilon}^{r_{\textrm{max}}}F_H\left(\frac{\gamma r^\alpha}{P_{T_S}}\right)\left(\frac{\left|\region-\pi r_c^2\right|}{\left|\region\right|}
 +\frac{\pi r_c^2}{\left|\region\right|}F_H\left(\frac{\gamma r^\alpha}{P_{T_S}}\right)\right)^{M-1}\fr dr.
 \end{align}
\else
For the cooperation-based protocol, the MGF of the interference at an arbitrarily located PU-Rx due to an independently and uniformly distributed SU inside an arbitrarily-shaped finite region is approximated by~\eqref{mgf_coop} shown at the top of next page.
\begin{figure*}[!t]
\normalsize
 \begin{align}\label{mgf_coop}
 \mgf \approx &\int_0^{\infty}\int_{\epsilon}^{r_{\textrm{max}}}\exp\left(-s P_{T} g r^{-\alpha}\right)F_H\left(\frac{\gamma r^\alpha}{P_{T_S}}\right)\left(\frac{\left|\region-\pi r_c^2\right|}{\left|\region\right|}
 +\frac{\pi r_c^2}{\left|\region\right|}F_H\left(\frac{\gamma r^\alpha}{P_{T_S}}\right)\right)^{M-1}\fr\fg dr dg  \nonumber\\
 &+1-\int_{\epsilon}^{r_{\textrm{max}}}F_H\left(\frac{\gamma r^\alpha}{P_{T_S}}\right)\left(\frac{\left|\region-\pi r_c^2\right|}{\left|\region\right|}
 +\frac{\pi r_c^2}{\left|\region\right|}F_H\left(\frac{\gamma r^\alpha}{P_{T_S}}\right)\right)^{M-1}\fr dr.
 \end{align}
\hrulefill
\vspace*{4pt}
\end{figure*}
\fi
\end{theorem}

\begin{corollary}\label{corollary_coop}
\ifCLASSOPTIONonecolumn
For the cooperation-based protocol, the $n$-th moment of the interference at an arbitrarily located PU-Rx due to an independently and uniformly distributed SU inside an arbitrarily-shaped finite region is approximated by
\begin{align}\label{moment_coop}
\moment\approx P_{T}^n \mathbb{E}_G\left\{G^n\right\}\mathbb{E}_R\left\{F_H\left(\frac{\gamma R^\alpha}{P_{T_S}}\right)\left(\frac{\left|\region-\pi r_c^2\right|}{\left|\region\right|}
 +\frac{\pi r_c^2}{\left|\region\right|}F_H\left(\frac{\gamma R^\alpha}{P_{T_S}}\right)\right)^{M-1}R^{-n\alpha}\right\}.
\end{align}
\else
For the cooperation-based protocol, the $n$-th moment of the interference at an arbitrarily located PU-Rx due to an independently and uniformly distributed SU inside an arbitrarily-shaped finite region is approximated by~\eqref{moment_coop} shown at the top of next page.
\begin{figure*}[!t]
\normalsize
\begin{align}\label{moment_coop}
\moment\approx P_{T}^n \mathbb{E}_G\left\{G^n\right\}\mathbb{E}_R\left\{F_H\left(\frac{\gamma R^\alpha}{P_{T_S}}\right)\left(\frac{\left|\region-\pi r_c^2\right|}{\left|\region\right|}
 +\frac{\pi r_c^2}{\left|\region\right|}F_H\left(\frac{\gamma R^\alpha}{P_{T_S}}\right)\right)^{M-1}R^{-n\alpha}\right\}.
\end{align}
\hrulefill
\vspace*{4pt}
\end{figure*}
\fi
\end{corollary}

Proof of Theorem~\ref{theorem_coop} and Corollary~\ref{theorem_coop}: See Appendix~\ref{app_coop}.

\begin{remark}
To the best of our knowledge, it is not possible to express~\eqref{mgf_coop} and~\eqref{moment_coop} in closed-form, even for the special cases of PU-Rx located at the center of a polygon or disk region. This is because~\eqref{mgf_coop} and~\eqref{moment_coop} contain one term, related to the RV $R$, which is raised to the power of $M-1$ ($M\geq2$) inside the expectation. Nevertheless,~\eqref{mgf_coop} and~\eqref{moment_coop} can easily be evaluated numerically. Also, if $r_c=0$, the term raised to the power of $M-1$ becomes one and~\eqref{mgf_coop} and~\eqref{moment_coop} reduce to~\eqref{mgf_single} and~\eqref{moment_single}.
\end{remark}

Summarizing, for an arbitrarily located PU-Rx inside an arbitrarily-shaped convex region and the different SU activity protocols, we can calculate (i) the MGF of the aggregate interference by substituting~\eqref{mgf_fix},~\eqref{mgf_single} and~\eqref{mgf_coop} into~\eqref{mgf-bpp}, (ii) the $n$-th cumulant of the aggregate interference by substituting~\eqref{moment_fix},~\eqref{moment_single} and~\eqref{moment_coop} into~\eqref{cu-bpp}, and (iii) the outage probability in the primary network by substituting~\eqref{mgf_fix},~\eqref{mgf_single} and~\eqref{mgf_coop} into~\eqref{outage_bpp1} and~\eqref{outage_bpp}.

\section{Average Number of Active Secondary Users}\label{section_averagenode}
The aggregate interference at the PU-Rx and the resulting outage probability are metrics to evaluate the performance of the primary network, which was the common focus of most prior studies on cognitive networks, e.g.,~\cite{Lee2012,Rabbachin2011,Mahmood2012,Vijayandran2012}. Ideally, the performance of the secondary network should also be evaluated. Furthermore, this should be done subject to a quality-of-service (QoS) constraint that the SINR of each active SU is maintained higher than a desired level. One way to do this analytically is to determine the SU throughput which can be defined as the expected spatial density of successful SU transmission and depends on (i) the number of active SUs over a certain region and (ii) whether each active SU is in outage or not, i.e., whether its SINR is above a certain threshold.

\redcom{The exact SINR distribution of an active SU (and consequently the SU throughput) in an arbitrarily-shaped underlay cognitive network is difficult to obtain because of two main reasons. \textit{Firstly}, for an arbitrarily-shaped region with a fixed number of nodes, the Binomial Point Process is non-stationary and an active SU's SINR is, therefore, location-dependent. Thus, the SINR of an active SU at a certain location (say origin) does not reflect the SINR of other active SUs. The difficulty in \textit{analytically averaging} the active SU's SINR over all possible locations in an arbitrarily-shaped region poses a significant challenge for analytical analysis. \textit{Secondly}, with the consideration of the different SU activity protocols, only the active SUs generate interference to other SUs and PU-Rx. This means that when accounting for the interference to a SU-Rx (which is the desired receiver for a certain SU), the distance between an interfering SU and PU-Rx is \textit{correlated} to the distance between this interfering SU and the SU-Rx. This distance correlation poses a second significant challenge for analytical analysis.}

In this work, in order to evaluate the performance of the secondary network in underlay cognitive networks, we study the average number of active SUs. The average number of active SUs is an analytically tractable performance metric, which can indirectly measure the SU throughput under certain conditions. For example, if the SINR threshold for SUs is not too high or each SU is sufficiently close to its desired receiver, it is possible that almost every active SU can transmit successfully. Under such conditions, \redcom{the average number of active SUs plays the dominant role in determining the aggregate throughput of SUs}\footnote{\redcom{We have confirmed this through extensive simulations, which are not included here due to space limitations.}}.

The main analytical result in this section is presented in Theorem~\ref{theo_meannode} below.
\begin{theorem}\label{theo_meannode}
For any SU activity protocol with independently and uniformly distributed SUs inside an arbitrarily-shaped finite region, the average number of active SUs is given by
\begin{align}\label{mean_node}
 \meannode=M\times\mu_{I}(0),
\end{align}
where $\mu_I(0)$ denotes the zero-th moment of the interference at an arbitrarily located PU-Rx from a random SU, which is dependent on the SU activity protocol.
\end{theorem}

Proof: See Appendix~\ref{app_mean_node}.

\begin{remark}
Theorem~\ref{theo_meannode} is valid for any SU activity protocol with i.u.d. node distribution and i.i.d. fading channels. For the protocols considered in this work, the value of $\mu_{I}(0)$ can be easily computed from~\eqref{moment_fix},~\eqref{moment_single} and~\eqref{moment_coop}, respectively.
\end{remark}

\begin{remark}
Intuitively, there is tradeoff between the primary network performance (i.e., in terms of the outage probability in the primary network) and the secondary network performance (i.e., in terms of the average number of active SUs). For example, increasing $r_f$ in the guard zone protocol or decreasing the activation threshold $\gamma$ in threshold-based and cooperation-based protocols can reduce the outage probability. However, this would decrease the number of active secondary users, which means the licensed spectrum is not efficiently reused. In this context,~\eqref{outage_bpp} and~\eqref{mean_node} provide an analytical means for evaluating this tradeoff in the performance of both the primary and secondary networks. In the next section, we will use the primary-secondary performance tradeoff as a systematic way to compare the performance of the different SU activity protocols.
\end{remark}

\section{Numerical Results}\label{section-result}
In this section, we present numerical results to investigate and compare the performance of the SU activity protocols. In order to validate the numerical results, we also present simulation results which are generated using MATLAB and are averaged over $1$ million simulation runs. For the simulation results, we use the following procedure to uniformly distribute the SUs inside an arbitrarily-shaped region~\cite{Martinez-2001}: (a) Generate a bounding box which is generally the minimal rectangle that can entirely enclose the polygon shape, (b) Randomly and uniformly generate a point in this bounding box, (c) Check whether this point is inside the required polygon, (d) Repeat steps (b) and (c) until the required number of nodes are obtained. Unless specified otherwise, the values of the main system parameters shown in Table~\ref{tb:3} are used. \redcom{All the distance, side length and radius values are in meters (m)}.
\begin{table}
\centering
\caption{Main System Parameter Values.}
\label{tb:3}
\begin{tabular}{|l|l|l|}\hline
Parameter & Symbol & Value \\
\hline
Transmit powers  & $P_{T_S}$, $P_{T_0}$, $P_T$ & $1$ \\ \hline
SINR threshold & $\beta$ &$0$ dB \\ \hline
Signal-to-noise ratio & $\rho_0$ & $20$ dB \\ \hline
Reference distance  & $r_0$ & $5$ \redcom{m} \\ \hline
Primary exclusion zone radius & $\epsilon$ & $1$ \\ \hline
Path-loss exponent & $\alpha$ & $2.5$ \\ \hline
Nakagami-$m$ fading parameters & $m_0=m_{g}=m_{h}$ & $3$ \\ \hline
\end{tabular}
\end{table}
%\begin{table}
%\centering
%\caption{Values of the Main System Parameters.}
%\label{tb:3}
%\begin{tabular}{|l|l|l||l|l|l|}\hline
%Parameter & Symbol & Value &Parameter & Symbol & Value \\
%\hline
%Number of secondary users & $M$ & 100 &Reference distance  & $r_0$ & $5$ \\ \hline
%Transmit powers  & $P_{T_S}$, $P_{T_0}$, $P_T$ & $1$ & Prohibition zone radius & $\epsilon$ & $1$ \\ \hline
%SINR threshold & $\beta$ &$0$ dB & Path-loss coefficient & $\alpha$ & $2.5$ \\ \hline
%Signal-to-noise ratio & $\rho_0$ & $20$ dB & Nakagami-$m$ fading parameters & $m_0=m_{g}=m_{h}$ & $3$ \\ \hline
%\end{tabular}
%\end{table}

\subsection{Validation of Cooperation-based Protocol Analysis}
  \ifCLASSOPTIONonecolumn
  \begin{figure}
        \centering
        \includegraphics[width=0.6  \textwidth]{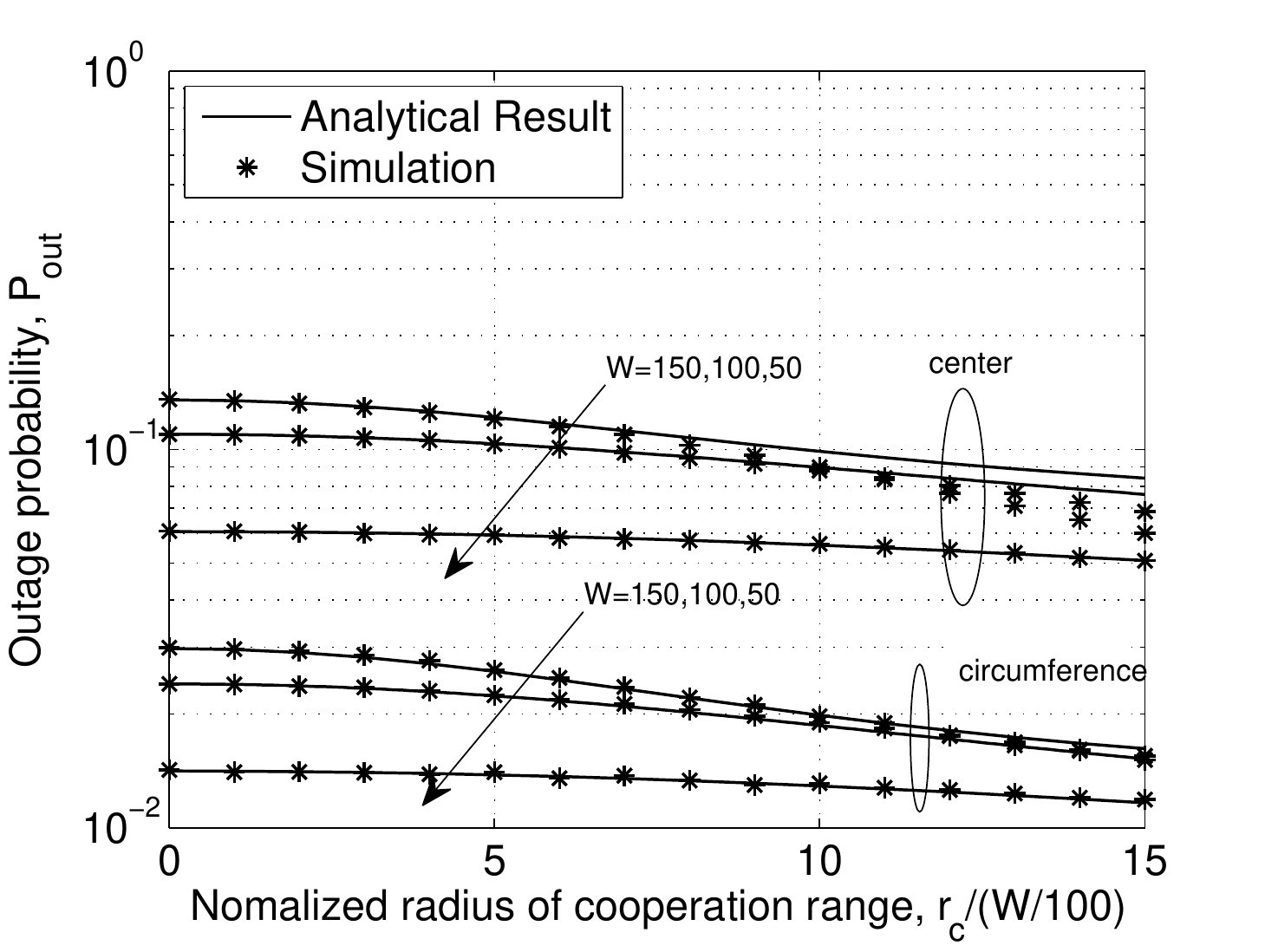}
        \caption{Outage probability, $\Pout$, versus the normalized radius of the cooperation range, $\frac{r_c}{W/100}$, for the cases that the PU-Rx is located at the center and circumference, respectively, of a disk region with different radius and number of SU pair values $(W,M)$ (i.e., $(150\;\mathrm{\redcom{m}},225)$, $(100\;\mathrm{\redcom{m}},100)$ and $(50\;\mathrm{\redcom{m}},25)$).}
        \label{coop_valid}
\end{figure}
\else
  \begin{figure}
        \centering
        \includegraphics[width=0.45  \textwidth]{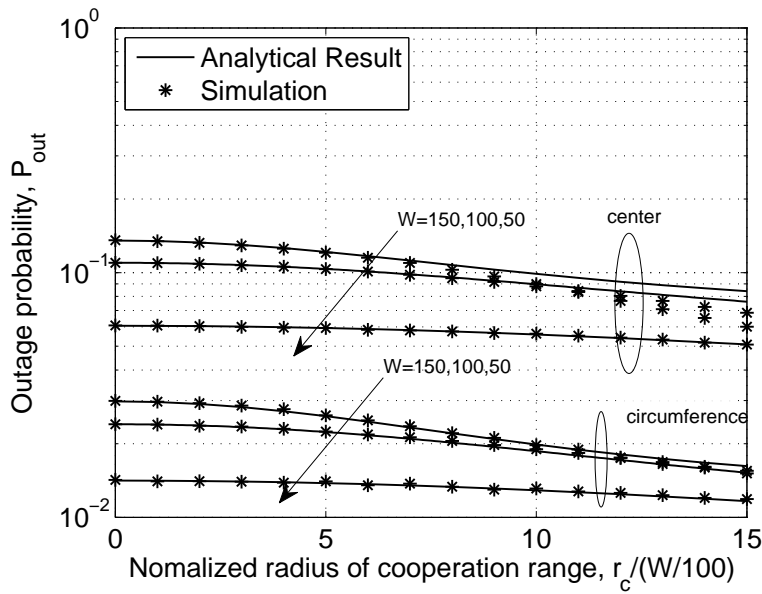}
        \caption{Outage probability, $\Pout$, versus the normalized radius of the cooperation range, $\frac{r_c}{W/100}$, for the cases that the PU-Rx is located at the center and circumference, respectively, of a disk region with different radius and number of SU pair values $(W,M)$ (i.e., $(150\;\mathrm{\redcom{m}},225)$, $(100\;\mathrm{\redcom{m}},100)$ and $(50\;\mathrm{\redcom{m}},25)$).}
        \label{coop_valid}
       % \vspace{-3mm}
\end{figure}
\fi

First, we investigate the accuracy of the cooperation-based protocol analysis given in Section~\ref{sec:coop:analysis}. Fig.~\ref{coop_valid} plots the outage probability, $\Pout$, versus the normalized radius of the cooperation range, $\frac{r_c}{W/100}$, when the PU-Rx is located at the center and circumference, respectively, of a disk region with different radius and number of SU pair values $(W,M)$ (i.e., $(150\;\mathrm{\redcom{m}},225)$, $(100\;\mathrm{\redcom{m}},100)$ and $(50\;\mathrm{\redcom{m}},25)$). The analytical result is plotted using Theorem~\ref{theorem_coop}, i.e., by substituting~\eqref{mgf_coop} into~\eqref{outage_bpp1} and~\eqref{outage_bpp}. The figure shows that the analytical results match closely with the simulation results when the cooperation range is relatively small compared to the size of the cognitive network region. For the disk case with the considered $(W,M)$ pair values, the analytical result is accurate even when the radius of the cooperation region is as large as $7\%$ of the radius of the disk region. This is in line with the assumption used for the analysis in Appendix~\ref{app_coop}. We can also see from the figure that the outage probability decreases as $r_c$ increases. This can be intuitively explained as follows. When the radius of the cooperation range $r_c$ increases, the number of cooperating SUs increases and the opportunity of being active for each SU decreases. This reduces the aggregate interference and improves the primary network's outage performance. \redcom{Also increasing the number of SUs increases the slope of the curves}.

\subsection{Moments of Aggregate Interference at the Primary Receiver}\label{app_cdf_subsection}
\ifCLASSOPTIONonecolumn
\else
\begin{table*}[t]
\centering
\caption{Validation of the 1st, 2nd and 3rd moment of the three SU activity protocols.}\label{tb:1}
\begin{tabular}{|c||c|c|c||c|c|c||c|c|c|} \hline
SU activity & \multicolumn{2}{c|}{$1$st moment}&{Percentage}&\multicolumn{2}{c|}{$2$nd moment}&{Percentage}&\multicolumn{2}{c|}{3rd moment}&{Percentage}\\
\cline{2-3}\cline{5-6}\cline{8-9}
 protocol & analytical & simulation &error ($\%$)& analytical & simulation &error ($\%$)& analytical & simulation&error ($\%$)\\\hline
  Guard zone&2.82E-3&2.82E-3&0.003&8.20E-6&8.20E-6&0.006&2.45E-8&2.45E-8&0.007\\ \hline
  Threshold-based&2.10E-3&2.10E-3&0.002&4.61E-6&4.61E-6&0.006&1.06E-8&1.05E-8&0.068\\ \hline
  Cooperation-based&1.83E-3&1.83E-3&0.063&3.47E-6&3.48E-6&0.29&6.88E-9&6.95E-9&1\\ \hline
\end{tabular}
\end{table*}
\fi
In this subsection, we investigate and compare the moments of the aggregate interference at the PU-Rx for the different SU activity protocols. We also illustrate the versatility of the proposed framework in being able to handle arbitrarily-shaped cognitive network regions.

We consider an arbitrarily-shaped cognitive network, as depicted in Fig. 1, with side lengths $S_1=S_2=\sqrt{3}\dist$, $S_3=\sqrt{7-3\sqrt{2}-\sqrt{6}}\dist$ and $S_4=\dist$ and interior angles $\theta_1=\pi/2$, $\theta_2=\pi/4$, $\theta_3=\pi-\arcsin\left(\frac{\sqrt{6}-\sqrt{2}}{2 S_3}\dist\right)$ and $\theta_4=\frac{\pi}{4}+\arcsin\left(\frac{\sqrt{6}-\sqrt{2}}{2 S_3}\dist\right)$. Without loss of generality, the origin is assumed to be at vertex $V_1$. The PU-Rx is located at coordinates $(\frac{\sqrt{3}\cos{\frac{3\pi}{8}}}{2\sin{\frac{11\pi}{24}}} \dist, \frac{\sqrt{3}\sin{\frac{3\pi}{8}}}{2\sin{\frac{11\pi}{24}}}\dist)$, which corresponds to the intersection point of the two diagonals of the $L=4$ sided arbitrarily-shaped cognitive network. The radius of the guard zone and the cooperation range are set to $r_f=30$ \redcom{m} and $r_c=8$ \redcom{m}, respectively. The activation threshold for both the threshold-based and the cooperation-based protocols is set to $\gamma=10^{-4}$ (linear scale).

\ifCLASSOPTIONonecolumn
Applying the algorithm in~\cite{Guo-2013}, the distance distribution function $\fr$ can be expressed as
\begin{small}
\begin{align}\label{pdf_distance}
\fr=\frac{1}{|\region|}\begin{cases}
              2\pi r, & {\epsilon\leq r<d_{S_3};} \\
           2\pi r -2r\arccos\left(\frac{d_{S_3}}{r}\right), & {d_{S_3}\leq r <d_{S_4};}\\
            2\pi r -2r\arccos\left(\frac{d_{S_3}}{r}\right)-2r\arccos\left(\frac{d_{S_4}}{r}\right), & {d_{S_4}\leq r <d_{V_4};}\\
              \frac{9}{4}\pi r-\theta_3 r -r\arccos\left(\frac{d_{S_3}}{r}\right)-r\arccos\left(\frac{d_{S_4}}{r}\right), & {d_{V_4}\leq r <d_{S_2};}\\
              \frac{9}{4}\pi r-\theta_3 r-2r\arccos\left(\frac{d_{S_2}}{r}\right)-r\arccos\left(\frac{d_{S_3}}{r}\right)-r\arccos\left(\frac{d_{S_4}}{r}\right), & {d_{S_2}\leq r <d_{V_3};}\\
               \frac{5}{4}\pi r-r\arccos\left(\frac{d_{S_2}}{r}\right)-r\arccos\left(\frac{d_{S_4}}{r}\right), & {d_{V_3}\leq r <d_{S_1};}\\
               \frac{5}{4}\pi r-2r\arccos\left(\frac{d_{S_1}}{r}\right)-r\arccos\left(\frac{d_{S_2}}{r}\right)-r\arccos\left(\frac{d_{S_4}}{r}\right), & {d_{S_1}\leq r <d_{V_1};}\\
                \frac{3}{4}\pi r-r\arccos\left(\frac{d_{S_1}}{r}\right)-r\arccos\left(\frac{d_{S_2}}{r}\right), & {d_{V_1}\leq r <d_{V_2};}\\
             \end{cases}
\end{align}
\end{small}
\else
Applying the algorithm in~\cite{Guo-2013}, the distance distribution function $\fr$ can be expressed as~\eqref{pdf_distance} shown at the top of next page,
\begin{figure*}[t]
\normalsize
\begin{align}\label{pdf_distance}
\fr=\frac{1}{|\region|}\begin{cases}
              2\pi r, & {\epsilon\leq r<d_{S_3};} \\
           2\pi r -2r\arccos\left(\frac{d_{S_3}}{r}\right), & {d_{S_3}\leq r <d_{S_4};}\\
            2\pi r -2r\arccos\left(\frac{d_{S_3}}{r}\right)-2r\arccos\left(\frac{d_{S_4}}{r}\right), & {d_{S_4}\leq r <d_{V_4};}\\
              \frac{9}{4}\pi r-\theta_3 r -r\arccos\left(\frac{d_{S_3}}{r}\right)-r\arccos\left(\frac{d_{S_4}}{r}\right), & {d_{V_4}\leq r <d_{S_2};}\\
              \frac{9}{4}\pi r-\theta_3 r-2r\arccos\left(\frac{d_{S_2}}{r}\right)-r\arccos\left(\frac{d_{S_3}}{r}\right)-r\arccos\left(\frac{d_{S_4}}{r}\right), & {d_{S_2}\leq r <d_{V_3};}\\
               \frac{5}{4}\pi r-r\arccos\left(\frac{d_{S_2}}{r}\right)-r\arccos\left(\frac{d_{S_4}}{r}\right), & {d_{V_3}\leq r <d_{S_1};}\\
               \frac{5}{4}\pi r-2r\arccos\left(\frac{d_{S_1}}{r}\right)-r\arccos\left(\frac{d_{S_2}}{r}\right)-r\arccos\left(\frac{d_{S_4}}{r}\right), & {d_{S_1}\leq r <d_{V_1};}\\
                \frac{3}{4}\pi r-r\arccos\left(\frac{d_{S_1}}{r}\right)-r\arccos\left(\frac{d_{S_2}}{r}\right), & {d_{V_1}\leq r <d_{V_2};}\\
             \end{cases}
\end{align}
\hrulefill
\vspace*{4pt}
\end{figure*}
\fi
\noindent where $d_{V_i}$ denotes the distance from PU-Rx to vertex $V_i$ ($i=1,2,3,4$), $d_{S_i}$ denotes the perpendicular distance from PU-Rx to side $S_i$ and the area $|\region|=\left(\frac{3}{2\sqrt{2}}+\frac{1}{2}\left(\sqrt{3}-\sqrt{\frac{3}{2}}\right)\right)\dist^2-\pi \epsilon^2$. Using the geometry, it can be easily shown that $d_{V_1}=\frac{\sqrt{3}}{2\sin{\frac{11\pi}{24}}} \dist$, $d_{V_2}=\frac{\sqrt{3}\sin{\frac{3\pi}{8}}}{\sin{\frac{11\pi}{24}}}\dist$, $d_{V_3}=\frac{\sqrt{6}\dist}{2\sin{\frac{3\pi}{8}}}-d_{V_1}$, $d_{V_4}=2\dist-d_{V_2}$, $d_{S_1}=\frac{\sqrt{3}\sin{\frac{3\pi}{8}}}{2\sin{\frac{11\pi}{24}}} \dist$, $d_{S_2}=\frac{\sin{\frac{13\pi}{24}}}{\sqrt{3}\dist}d_{V_2}d_{V_3}$, $d_{S_3}=\frac{\sin{\frac{11\pi}{24}}}{S_3}d_{V_3}d_{V_4}$, and $d_{S_4}= \frac{\sqrt{3}\cos{\frac{3\pi}{8}}}{2\sin{\frac{11\pi}{24}}}\dist$.
\ifCLASSOPTIONonecolumn
\begin{table*}[t]
\centering
\caption{Validation of the 1st, 2nd and 3rd moment of the three SU activity protocols.}\label{tb:1}
\begin{tabular}{|c||c|c|c||c|c|c||c|c|c|} \hline
SU activity & \multicolumn{2}{c|}{$1$st moment}&{Percentage}&\multicolumn{2}{c|}{$2$nd moment}&{Percentage}&\multicolumn{2}{c|}{3rd moment}&{Percentage}\\
\cline{2-3}\cline{5-6}\cline{8-9}
 protocol & analytical & simulation &error ($\%$)& analytical & simulation &error ($\%$)& analytical & simulation&error ($\%$)\\\hline
  Guard zone&2.82E-3&2.82E-3&0.003&8.20E-6&8.20E-6&0.006&2.45E-8&2.45E-8&0.007\\ \hline
  Threshold-based&2.10E-3&2.10E-3&0.002&4.61E-6&4.61E-6&0.006&1.06E-8&1.05E-8&0.068\\ \hline
  Cooperation-based&1.83E-3&1.83E-3&0.063&3.47E-6&3.48E-6&0.29&6.88E-9&6.95E-9&1\\ \hline
\end{tabular}
\end{table*}
\else
\fi

Substituting~\eqref{pdf_distance} in Corollaries 1$-$3, we can obtain the analytical $n$-th moment results. Table~\ref{tb:1} shows the 1st, 2nd and 3rd moment of the aggregate interference for the three SU activity protocols with $\dist=150$ \redcom{m}. The simulation results are an excellent match with the analytical results, which confirms the accuracy of the results in Corollaries 1$-$3. We can see from Table~\ref{tb:1} that for the considered case, the proposed cooperation-based protocol has the smallest values of the moments, i.e., it results in the smallest aggregate interference. The next best is the threshold-based protocol, followed by the guard zone protocol.
\subsection{Outage Probability at the Primary Receiver}\label{app_cdf_subsection2}
In this subsection, we investigate and compare the outage probability at the PU-Rx for the different SU activity protocols.
\ifCLASSOPTIONonecolumn
\begin{figure}
\centering
        \includegraphics[width=0.6  \textwidth]{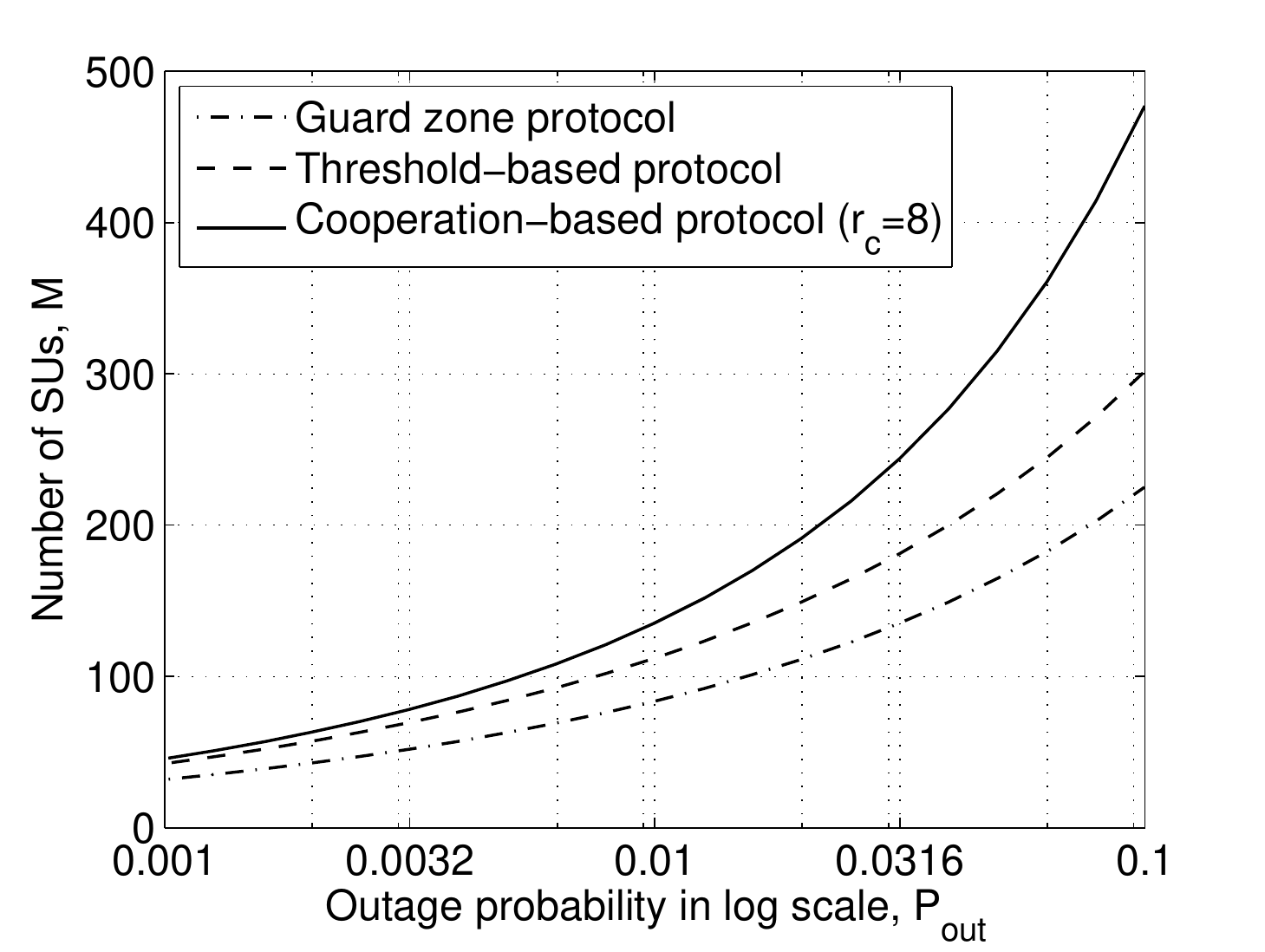}
        \caption{Number of SUs, $M$, versus the outage probability, $P_{\textrm{out}}$, for the scenario defined in Section VI-B with $\dist=150$ \redcom{m}.}
        \label{outage_arb}
\end{figure}
\else
\begin{figure}
\centering
        \includegraphics[width=0.45  \textwidth]{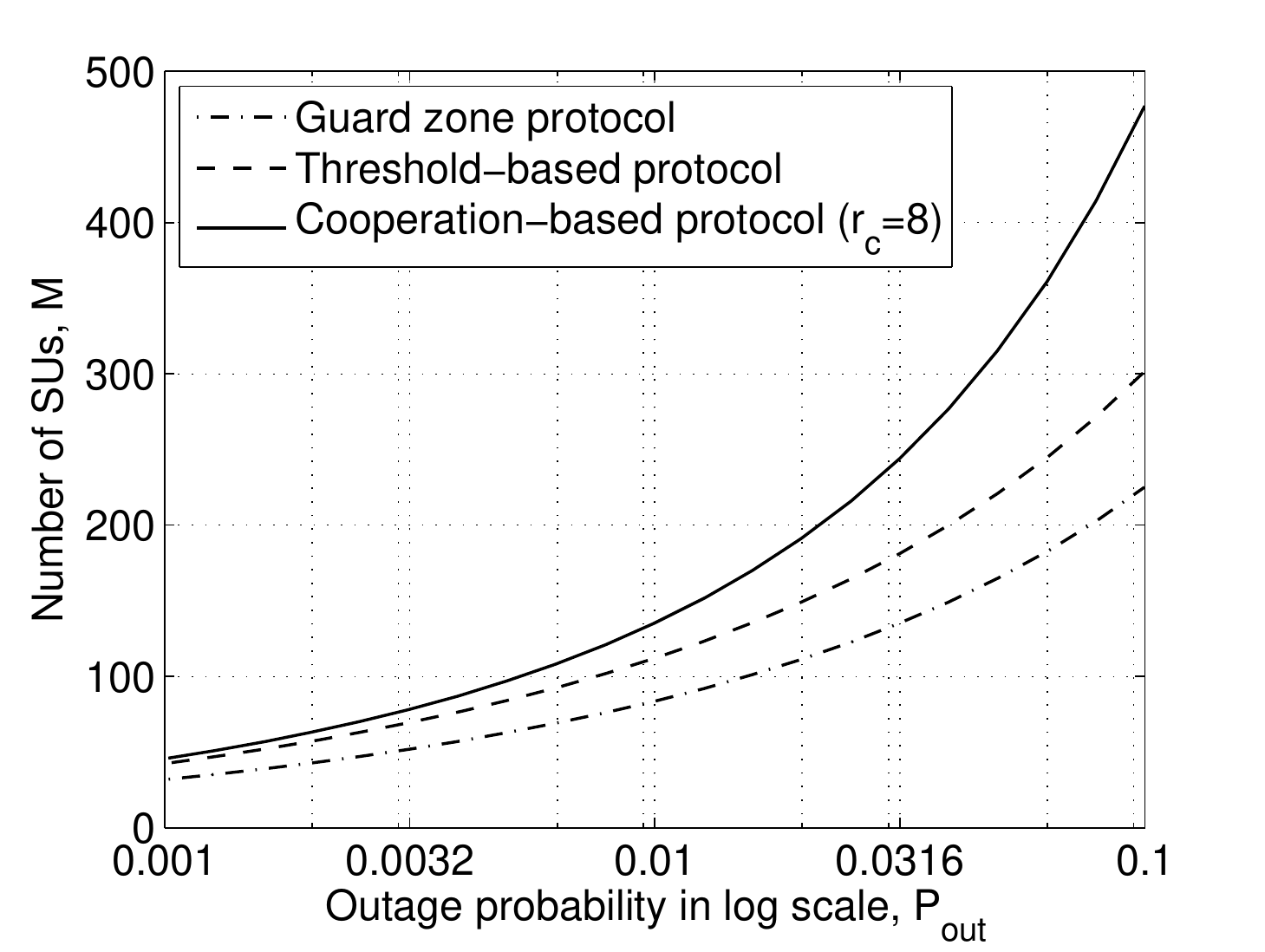}
        \caption{Number of SUs, $M$, versus the outage probability, $P_{\textrm{out}}$, for the scenario defined in Section VI-B with $\dist=150$ \redcom{m}.}
        \label{outage_arb}
         %\vspace{-3mm}
\end{figure}
\fi
Fig.~\ref{outage_arb} plots the number of SUs $M$ versus the outage probability, $P_{\textrm{out}}$, for the different SU activity protocols and the arbitrarily-shaped cognitive network region considered in Section~\ref{app_cdf_subsection} with $\dist=150$ \redcom{m}. The analytical results are plotted by substituting~\eqref{pdf_distance} in Theorems 1$-$3 and then substituting in~\eqref{outage_bpp1} and~\eqref{outage_bpp}. As illustrated in this figure, for the considered case, the cooperation-based performs the best while the guard zone protocol performs the worst. For example, for $M=100$ SUs inside the arbitrarily-shaped region, the cooperation-based protocol can achieve the best QoS of $\Pout=5.33*10^{-3}$ at the PU-Rx. However, the threshold-based protocol and guard zone protocol can only achieve $\Pout=7.59*10^{-3}$ and $\Pout=1.54*10^{-2}$, respectively.

It must be noted that such \redcom{an analysis} has two main issues. Firstly, for a given shape of the network region and a given location of the PU-Rx, the outage at the PU-Rx strongly relies on the protocol system parameters (i.e., the guard zone range $r_f$ for the guard zone protocol, activation threshold $\gamma$ for both of the threshold-based and cooperation-based protocols) and different values of the protocol system parameters can lead to a different performance ordering of the SU activity protocols. Secondly, it focuses on the performance in the primary network only. These aspects are addressed in the next Section~\ref{result_compare}.
   %We can see that the analytical results match perfectly with the simulation results, which verifies the results in Theorems 1$-$3.

\subsection{Comparison of Secondary User Activity Protocols}\label{result_compare}
In this section, we investigate and compare the SU activity protocols in terms of their effect on the primary network (i.e., the outage probability) and the secondary network (i.e., the average number of active SUs).
\ifCLASSOPTIONonecolumn
 \begin{figure}
\centering
\subfigure[Different path-loss exponent $\alpha=2,3,6$. ]{\label{compare1}\includegraphics[width=0.495\textwidth]{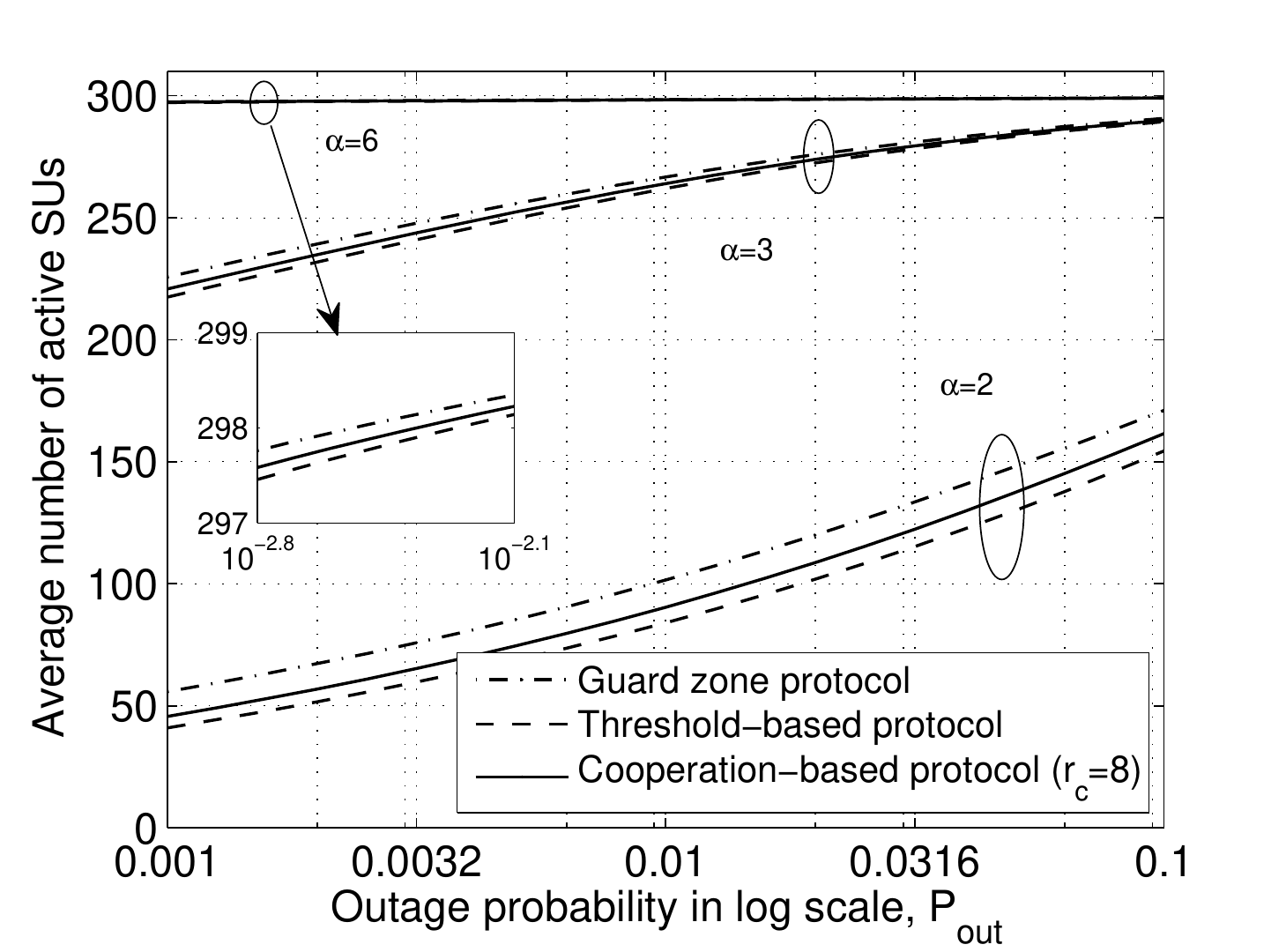}}
%\mbox{\hspace{0.5cm}}
\subfigure[Different Nakagami-$m$ fading $m=1,3,5$. ]{\label{compare2}\includegraphics[width=0.495\textwidth]{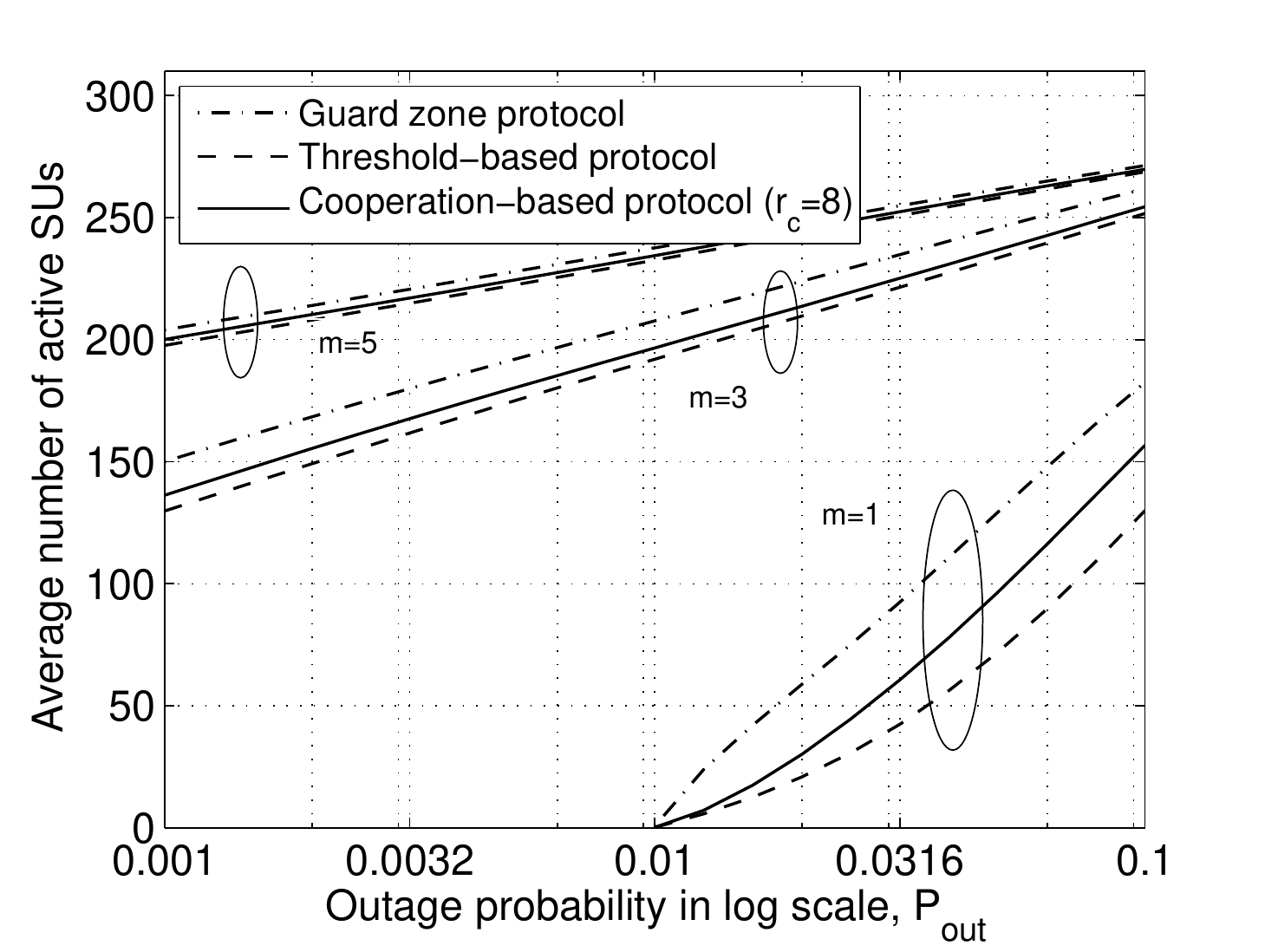}}
\caption{Average number of active SUs, $\meannode$, versus the outage probability, $P_{\textrm{out}}$, when the PU-Rx is located at the intersection point of two diagonals inside the arbitrarily-shaped finite region specified in Section VI-B with $\dist=150$ \redcom{m}.}
\label{compare}
\end{figure}

\else
 \begin{figure}
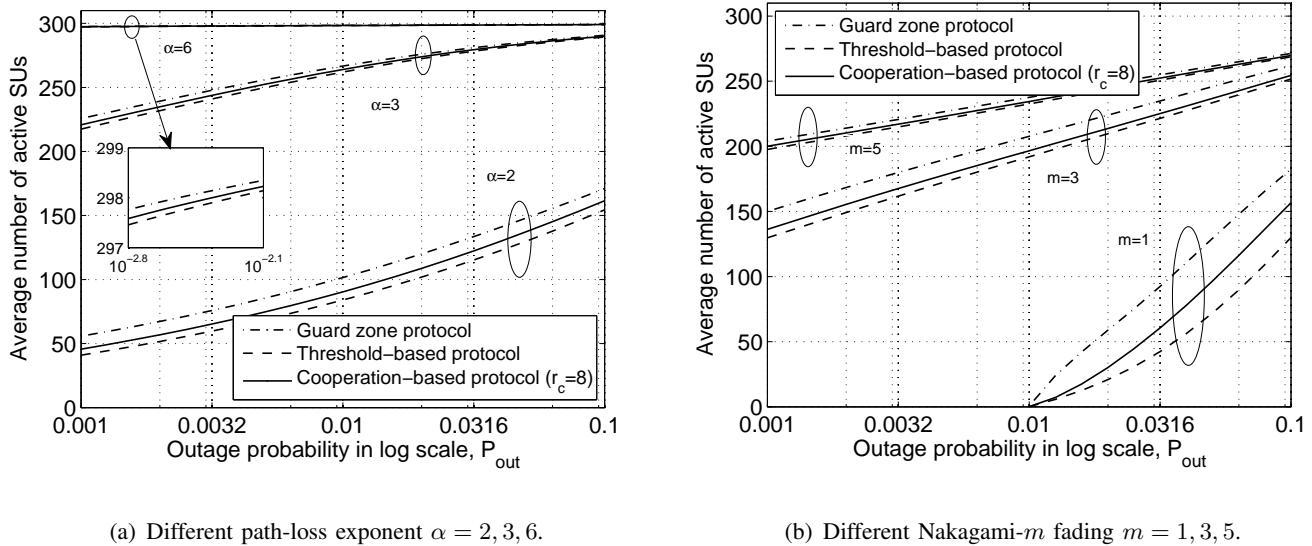

 \vspace{-4mm}
\centering
\subfigure[Different path-loss exponent $\alpha=2,3,6$. ]{\label{compare1}\includegraphics[width=0.46\textwidth]{fig/figure5_a1}}
%\mbox{\hspace{0.5cm}}
\subfigure[Different Nakagami-$m$ fading $m=1,3,5$. ]{\label{compare2}\includegraphics[width=0.46\textwidth]{fig/figure5_b}}
\caption{Average number of active SUs, $\meannode$, versus the outage probability, $P_{\textrm{out}}$, when the PU-Rx is located at the intersection point of two diagonals inside the arbitrarily-shaped finite region specified in Section VI-B with $\dist=150$ \redcom{m}.}
\label{compare}
 \vspace{-3mm}
\end{figure}
\fi

Fig.~\ref{compare} plots the average number of active SUs, $\meannode$, versus the outage probability, $P_{\textrm{out}}$, when the PU-Rx is located at the intersection point of the two diagonals inside the arbitrarily-shaped finite region specified in Section VI-B with $\dist=150$ \redcom{m}. The results are shown for different values of path-loss exponent $\alpha=2,3,6$ (Fig.~\ref{compare1}) and Nakagami-$m$ fading $m=1,3,5$ (Fig.~\ref{compare2}), respectively. The following approach is adopted to work out these curves for the SU activity protocols:
 \begin{enumerate}
   \item For each outage probability value, using~\eqref{outage_bpp},~\eqref{outage_bpp1} and Theorems $1-3$, we can find the value of $r_f$ for guard zone protocol and $\gamma$ for threshold-based protocol and cooperation-based protocol, respectively;
   \item We substitute the value $r_f$ and $\gamma$ into~\eqref{mean_node} and Corollaries $1-3$ to obtain the average number of active SUs for each protocol.
 \end{enumerate}

Fig.~\ref{compare} shows that for each protocol, the active number of SUs is higher if the fading is less severe (i.e., $m$ has higher value) or if the path-loss exponent has a higher value. For the same outage probability, \redcom{it is desirable} to have a larger average number of active SUs. From Fig.~\ref{compare}, we can see that the guard zone protocol \redcom{supports the largest number of active SUs}, followed by the cooperation-based protocol and the threshold-based protocol. For example, in order to achieve a QoS of $P_{\textrm{out}}=0.01$ at the PU-Rx with $m=3$ and $\alpha=2.5$, only $192$ SUs are active on average under the threshold-based protocol. However, the cooperation-based protocol and guard zone protocol can have $197$ and $207$ active SUs, respectively. This ordering stays the same for the different values of $m$ and $\alpha$.

This ordering can be intuitively explained as follows. The guard zone protocol \redcom{supports the largest number of active SUs} because it enables the SUs to determine their ``average'' impact on the PU-Rx from long-term sensing signal monitoring. The outage probability is usually caused by the SUs that are close to the PU-Rx. By forming a proper guard zone region around the PU-Rx, the interference from the nearby SUs is avoided and on average more SUs are allowed to transmit. The threshold-based protocol \redcom{supports the smallest number of active SUs} due to the effect of uncorrelated fading between sensing and transmitting channels. For example, the fading on the sensing channel may be severe but the fading on the SU transmitting channel may be weak. Thus, the SU may receive a weak signal on the sensing channel and decide to transmit, which may generate severe interference for the PU-Rx. The cooperation-based protocol helps to minimize the effect of uncorrelated fading channels by utilizing the local information exchange among SUs. Therefore, the cooperation-based protocol allows a higher activation threshold while still meeting the same outage value. This results in the larger number of active SUs, compared to the threshold-based protocol. Although the guard zone protocol \redcom{supports the largest number of active SUs}, the implementation of this protocol is only possible in scenarios that allow a long-term sensing signal monitoring before SUs can transmit. On the other hand, the threshold-based and cooperation-based protocols only rely on the short-term signal monitoring. \redcom{It must be noted that a more comprehensive comparison requires a rigorous study of the tradeoff between performance and implementation complexity in the scenario of interest, which is outside the scope of this paper.}

\section{Conclusions}\label{conclusion}
In this paper, we have proposed a framework to analyze the performance of arbitrarily-shaped underlay cognitive networks with different SU activity protocols. The framework depends on the MGF of the interference due to a random SU and allows closed-form computation of the outage probability in the primary network as well as the average number of active SUs in the secondary network. We have also applied cooperation in the context of the underlay cognitive network to come up with a cooperation-based SU activity protocol, which includes the existing threshold-based protocol as a special case. \redcom{We studied the average number of active SUs for the different SU activity protocols, subject to a given outage probability constraint at the PU} and we used it as an analytical approach to compare the performance of different SU activity protocols in the underlay cognitive networks. Our results showed that, in the short-term sensing signal monitoring scenarios, the cooperation-based protocol improves the networks' performance compared to the threshold-based protocol. The proposed framework is especially relevant for emerging ultra-dense small cell deployment scenarios, where network regions can be arbitrarily-shaped. Future work can consider the generalization of cooperation-based protocol with multiple activation thresholds, which is outside the scope of the present work. It can also analyse the \redcom{throughput} performance of the SUs subject to a quality-of-service constraint.

\appendices

\section{Proof of Theorems and Corollaries}\label{sec:wholeproof}
In this appendix, we derive the moment generating function of the interference from a random SU and the corresponding $n$-th moment for the SU activity protocols.

From~\eqref{intf_guard},~\eqref{single} and~\eqref{intf_coop}, we can see that whether a random SU generates interference or not is strongly dependent on the SU's random distance to the PU-Rx, $R$. The conditional probability mass function (PMF) of the interference from a SU is given by
\begin{align}\label{general_pmf}
\Pr (I=\mathds{I}|R)=\begin{cases}
              P_{\textrm{active}|R}, & {\mathds{I}=P_T G R^{-\alpha};} \\
           1-P_{\textrm{active}|R}, & {\mathds{I}=0;}
             \end{cases}
\end{align}
\noindent where $P_{\textrm{active}|R}$ represents the conditional probability that a SU is active, which is conditioned on its random distance $R$ to the PU-Rx.

Using~\eqref{general_pmf}, we can express the MGF of the interference from a random SU (defined below~\eqref{mgf-bpp}) as
\ifCLASSOPTIONonecolumn
\begin{align}\label{general_mgf}
\mgf =&\mathbb{E}_I\left\{\exp(-sI)\right\}\nonumber\\
=&\mathbb{E}_{G}\left\{\mathbb{E}_{R}\left\{\Pr (I=\mathds{I}|R)\exp\left(-s\mathds{I}\right)\right\}\right\} \nonumber\\
=&\mathbb{E}_{G}\left\{\mathbb{E}_{R}\left\{P_{\textrm{active}|R}\exp\left(-sP_T G R^{-\alpha}\right\}\right\}\right\}+\mathbb{E}_{G}\left\{\mathbb{E}_{R}\left\{\left(1-P_{\textrm{active}|R}\right)\exp\left(-s\times 0\right\}\right\}\right\} \nonumber\\
=&\mathbb{E}_{G,R}\left\{P_{\textrm{active}|R}\exp\left(-sP_T G R^{-\alpha}\right)\right\}+1-\mathbb{E}_{R}\left\{P_{\textrm{active}|R}\right\}.
\end{align}
\else
\begin{align}\label{general_mgf}
\mgf =&\mathbb{E}_I\left\{\exp(-sI)\right\}\nonumber\\
=&\mathbb{E}_{G}\left\{\mathbb{E}_{R}\left\{\Pr (I=\mathds{I}|R)\exp\left(-s\mathds{I}\right)\right\}\right\} \nonumber\\
=&\mathbb{E}_{G}\left\{\mathbb{E}_{R}\left\{P_{\textrm{active}|R}\exp\left(-sP_T G R^{-\alpha}\right\}\right\}\right\}\nonumber\\
&+\mathbb{E}_{G}\left\{\mathbb{E}_{R}\left\{\left(1-P_{\textrm{active}|R}\right)\exp\left(-s\times 0\right\}\right\}\right\} \nonumber\\
=&\mathbb{E}_{G,R}\left\{P_{\textrm{active}|R}\exp\left(-sP_T G R^{-\alpha}\right)\right\}+1\nonumber\\
&-\mathbb{E}_{R}\left\{P_{\textrm{active}|R}\right\}.
\end{align}
\fi

Substituting~\eqref{general_mgf} into~\eqref{moment_relation}, the $n$-th moment of the interference from a random SU is
\ifCLASSOPTIONonecolumn
\begin{align}\label{general_moment}
\moment=&(-1)^n \mathbb{E}_{G,R}\left\{P_{\textrm{active}|R}\left(-1\right)^n P_T^n G^n R^{-n\alpha}\exp\left(-sP_T G R^{-\alpha}\right)\right\}|_{s=0}\nonumber\\
=& P_T^n \mathbb{E}_{G}\left\{G^n\right\}\mathbb{E}_{R}\left\{P_{\textrm{active}|R}R^{-n\alpha}\right\}.
\end{align}
\else
\begin{align}\label{general_moment}
\moment=&(-1)^n \mathbb{E}_{G,R}\left\{P_{\textrm{active}|R}\left(-1\right)^n P_T^n G^n R^{-n\alpha}\times\right.\nonumber\\
&\left.\exp\left(-sP_T G R^{-\alpha}\right)\right\}|_{s=0}\nonumber\\
= &P_T^n \mathbb{E}_{G}\left\{G^n\right\}\mathbb{E}_{R}\left\{P_{\textrm{active}|R}R^{-n\alpha}\right\}.
\end{align}
\fi

We now define and use the value of the conditional probability $P_{\textrm{active}|R}$ for different SU activity protocol to derive the main analytical results in the paper.

\subsection{Proof of Theorem~\ref{theorem_guard} and Corollary~\ref{theorem_guard} }\label{app_guard}
\begin{proof}
For the guard zone protocol,~\eqref{intf_guard} shows that the interference from a random SU is given by $P_{T}G R^{-\alpha}\textbf{1}_{\left(R>r_f\right)}$. Thus, the conditional probability that a SU is active can be expressed as
\begin{align}\label{active_guardzone}
P_{\textrm{active}|R}=\begin{cases}
              1, & {R> r_f;} \\
            0, & {R\leq r_f;}
             \end{cases}
\end{align}
Substituting~\eqref{active_guardzone} into~\eqref{general_mgf} and~\eqref{general_moment}, the MGF and $n$-th moment of the interference from a random SU for the guard zone protocol are respectively given by
\ifCLASSOPTIONonecolumn
\begin{align}
\mgf =&\int_0^{\infty}\left(\int_{\epsilon}^{r_f}0\times\exp\left(-s P_T g r^{-\alpha}\right)\fr dr +\int_{r_f}^{r_{\textrm{max}}}1\times\exp\left(-s P_T g r^{-\alpha}\right)\fr dr\right)\fg dg \nonumber\\
&+1-\left(\int_{\epsilon}^{r_f}0\times\fr dr+\int_{r_f}^{r_{\textrm{max}}}1\times\fr dr\right) \nonumber\\
=&\int_0^{\infty}\int_{r_f}^{r_{\textrm{max}}}\exp\left(-s P_{T}g r^{-\alpha}\right)\fr \fg dr dg +F_R(r_f),
\end{align}
\else
\begin{align}
\mgf =&\int_0^{\infty}\left(\int_{\epsilon}^{r_f}0\times\exp\left(-s P_T g r^{-\alpha}\right)\fr dr \right.\nonumber\\
&\left.+\int_{r_f}^{r_{\textrm{max}}}1\times\exp\left(-s P_T g r^{-\alpha}\right)\fr dr\right)\fg dg \nonumber\\
&+1-\left(\int_{\epsilon}^{r_f}0\times\fr dr+\int_{r_f}^{r_{\textrm{max}}}1\times\fr dr\right) \nonumber\\
=&\int_0^{\infty}\int_{r_f}^{r_{\textrm{max}}}\exp\left(-s P_{T}g r^{-\alpha}\right)\fr \fg dr dg +F_R(r_f),
\end{align}
\fi
and
\ifCLASSOPTIONonecolumn
\begin{align}
\moment=& P_T^n \mathbb{E}_{G}\left\{G^n\right\}\left(\int_\epsilon^{r_f}0\times R^{-n\alpha}\fr dr+\int_{r_f}^{r_{\textrm{max}}}1\times R^{-n\alpha}\fr dr\right)\nonumber \\
=&P_T^n \mathbb{E}_{G}\left\{G^n\right\}\int_{r_f}^{r_{\textrm{max}}} R^{-n\alpha}\fr dr.
\end{align}
\else
\begin{align}
\moment=& P_T^n \mathbb{E}_{G}\left\{G^n\right\}\left(\int_\epsilon^{r_f}0\times R^{-n\alpha}\fr dr\right.\nonumber\\
&\left.+\int_{r_f}^{r_{\textrm{max}}}1\times R^{-n\alpha}\fr dr\right)\nonumber \\
=&P_T^n \mathbb{E}_{G}\left\{G^n\right\}\int_{r_f}^{r_{\textrm{max}}} R^{-n\alpha}\fr dr.
\end{align}
\fi

Hence, we arrive at the results in Theorem~\ref{theorem_guard} and Corollary~\ref{corollary_guard}.
\end{proof}
\subsection{Proof of Theorem~\ref{theorem_single} and Corollary~\ref{theorem_single}}\label{app_single}
\begin{proof}
For the threshold-based protocol,~\eqref{single} shows that the interference from a random SU is given by $P_{T}G R^{-\alpha}\textbf{1}_{\left(P_{T_S}H R^{-\alpha}\leq\gamma\right)}$, i.e., the SU generates interference as long as $H\leq \frac{\gamma R^\alpha}{P_{T_S}}$ when the distance to PU-Rx is given. Thus, the conditional probability that a SU is active can be written as
\begin{align}\label{active_single}
P_{\textrm{active}|R}=\int_{0}^{\frac{\gamma R^\alpha}{P_{T_S}}}\fh dh=F_H\left(\frac{\gamma R^\alpha}{P_{T_S}}\right).
\end{align}

Substituting~\eqref{active_single} into~\eqref{general_mgf} and~\eqref{general_moment}, we can express the MGF and $\moment$ of the interference from a random SU as
\ifCLASSOPTIONonecolumn
\begin{align}
\mgf =& \int_0^{\infty}\int_{\epsilon}^{r_{\textrm{max}}}\exp\left(-s P_T g r^{-\alpha}\right)F_H\left(\frac{\gamma r^\alpha}{P_{T_S}}\right)\fr\fg dr dg+1-\int_0^{\infty}F_H\left(\frac{\gamma r^\alpha}{P_{T_S}}\right)\fr dr,
\end{align}
\else
\begin{small}
\begin{align}
\mgf =& \int_0^{\infty}\int_{\epsilon}^{r_{\textrm{max}}}\exp\left(-s P_T g r^{-\alpha}\right)F_H\left(\frac{\gamma r^\alpha}{P_{T_S}}\right)\fr\fg dr dg\nonumber\\
&+1-\int_0^{\infty}F_H\left(\frac{\gamma r^\alpha}{P_{T_S}}\right)\fr dr,
\end{align}
\end{small}
\fi
and
\begin{align}
\moment= P_T^n \mathbb{E}_{G}\left\{G^n\right\}\mathbb{E}_{R}\left\{F_H\left(\frac{\gamma R^\alpha}{P_{T_S}}\right)R^{-n\alpha}\right\}.
\end{align}

Hence, we arrive at the results in Theorem~\ref{theorem_single} and Corollary~\ref{theorem_single}.
\end{proof}
\subsection{Proof of Theorem~\ref{theorem_coop} and Corollary~\ref{theorem_coop}}\label{app_coop}
\begin{proof}
For the cooperation-based protocol, consider a typical SU node A, cooperating with node B, as shown in Fig~\ref{coop}. Given the position of node A, let $p_{nt}$ represents the conditional probability that node B leads to node A deciding not to transmit. This event occurs when node B falls into the cooperative region $\mathcal{D}_A(r_c)$ around the typical user A and the received signal at node B is greater than the activation threshold $\gamma$. Using this fact, we can express $p_{nt}$ as
\ifCLASSOPTIONonecolumn
 \begin{align}\label{prob}
 p_{nt}=\int_{\mathcal{D}_A(r_c)\times\region}\int_{\frac{\gamma r_B^\alpha}{P_{T_S}}}^{\infty}f_{H_B}\left(h_B\right)dh_B dx_B dy_B=\int_{\mathcal{D}_A(r_c)\times\region}\left(1-F_{H_B}\left(\frac{\gamma r_B^\alpha}{P_{T_S}}\right)\right)dx_B dy_B,
 \end{align}
\else
 \begin{align}\label{prob}
 p_{nt}=&\int_{\mathcal{D}_A(r_c)\times\region}\int_{\frac{\gamma r_B^\alpha}{P_{T_S}}}^{\infty}f_{H_B}\left(h_B\right)dh_B dx_B dy_B\nonumber\\
 =&\int_{\mathcal{D}_A(r_c)\times\region}\left(1-F_{H_B}\left(\frac{\gamma r_B^\alpha}{P_{T_S}}\right)\right)dx_B dy_B,
 \end{align}
\fi
\noindent where $r_B=\sqrt{x_B^2+y_B^2}$ is the distance from node B to the PU-Rx\footnote{The location of PU-Rx is assumed to be at the origin.}, $\left(x_B, y_B\right)$ is the coordinate of node B, $f_{H_B}(h_B)$ represents the fading power distribution on the user B's sensing channel and $\int_{\mathcal{D}_A(r_c)\times\region}$ denotes the integration over the overlap region between $\region$ and $\mathcal{D}_A(r_c)$.

We can see from~\eqref{prob} that $p_{nt}$ is a function of the location of node A. Consequently, the integration in~\eqref{prob} is very complicated to evaluate in closed-form. In order to simplify the analysis, we assume that:
 \begin{itemize}
  \item the cooperation range $r_c$ is small compared to the size of the cognitive network region;
   \item the SUs within the cooperation range experience the same path-loss;
    \item the effect from the boundary is neglected so that the overlap region is the same as the cooperative region irrespective of location of node A.
    \end{itemize}
Thus, we can approximate~\eqref{prob} as
 \ifCLASSOPTIONonecolumn
  \begin{align}\label{prob2}
 p_{nt}\approx \left(1-F_{H_B}\left(\frac{\gamma R_A^\alpha}{P_{T_S}}\right)\right)\int_{\mathcal{D}_A(r_c)}dx_B dy_B  =\frac{\pi r_c^2}{\left|\region\right|}
 -\frac{\pi r_c^2}{\left|\region\right|}F_{H_B}\left(\frac{\gamma R_A^\alpha}{P_{T_S}}\right),
 \end{align}
\else
  \begin{align}\label{prob2}
 p_{nt}\approx &\left(1-F_{H_B}\left(\frac{\gamma R_A^\alpha}{P_{T_S}}\right)\right)\int_{\mathcal{D}_A(r_c)}dx_B dy_B \nonumber\\
  =&\frac{\pi r_c^2}{\left|\region\right|}
 -\frac{\pi r_c^2}{\left|\region\right|}F_{H_B}\left(\frac{\gamma R_A^\alpha}{P_{T_S}}\right),
 \end{align}
\fi
where $R_A$ is the distance from node A to the PU-Rx.

The complement of the probability $p_{nt}$, denoted by $1-p_{nt}$, is known as the probability that node B causes node A to transmit. In addition to node A, there are a total number of $M-1$ SUs which are independently distributed inside the network region. Consequently, the conditional probability that $M-1$ nodes can make node A to be active is $\left(1-p_{nt}\right)^{M-1}$, where $p_{nt}$ is given by~\eqref{prob2}.

In order for node A to transmit, both the received signal powers at the SUs inside $\mathcal{D}_A(r_c)$ and the received signal power by node A on the sensing channel must be less than the activation threshold $\gamma$. Since nodes are identically distributed, we can drop the subscript $A$ in $R_A$ and B in $F_{H_B}(\cdot)$. Thus, the conditional probability of a random SU being active can be expressed as
  \ifCLASSOPTIONonecolumn
\begin{align}\label{active_coop}
P_{\textrm{active}|R}\approx  \left(1-p_{nt}\right)^{M-1}\int_{0}^{\frac{\gamma R^\alpha}{P_{T_S}}}\fh dh=\left(\frac{\left|\region-\pi r_c^2\right|}{\left|\region\right|}
 +\frac{\pi r_c^2}{\left|\region\right|}F_H\left(\frac{\gamma r^\alpha}{P_{T_S}}\right)\right)^{M-1}F_H\left(\frac{\gamma r^\alpha}{P_{T_S}}\right).
\end{align}
\else
\begin{align}\label{active_coop}
P_{\textrm{active}|R}\approx & \left(1-p_{nt}\right)^{M-1}\int_{0}^{\frac{\gamma R^\alpha}{P_{T_S}}}\fh dh\nonumber\\
=&\left(\frac{\left|\region-\pi r_c^2\right|}{\left|\region\right|}
 +\frac{\pi r_c^2}{\left|\region\right|}F_H\left(\frac{\gamma r^\alpha}{P_{T_S}}\right)\right)^{M-1}F_H\left(\frac{\gamma r^\alpha}{P_{T_S}}\right).
\end{align}
\fi

\ifCLASSOPTIONonecolumn
Then the MGF and $n$-th for the interference from a random SU can be obtained by substituting~\eqref{active_coop} into~\eqref{general_mgf} and~\eqref{general_moment} respectively, which can be expressed as
 \begin{align}\label{proof_coop_eq}
 \mgf \approx & \int_0^{\infty}\int_{\epsilon}^{r_{\textrm{max}}}\exp\left(-s P_{T} g r^{-\alpha}\right)F_H\left(\frac{\gamma r^\alpha}{P_{T_S}}\right)\left(\frac{\left|\region-\pi r_c^2\right|}{\left|\region\right|}
 +\frac{\pi r_c^2}{\left|\region\right|}F_H\left(\frac{\gamma r^\alpha}{P_{T_S}}\right)\right)^{M-1}\fr\fg dr dg  \nonumber\\
 &+1-\int_{\epsilon}^{r_{\textrm{max}}}F_H\left(\frac{\gamma r^\alpha}{P_{T_S}}\right)\left(\frac{\left|\region-\pi r_c^2\right|}{\left|\region\right|}
 +\frac{\pi r_c^2}{\left|\region\right|}F_H\left(\frac{\gamma r^\alpha}{P_{T_S}}\right)\right)^{M-1}\fr dr,
 \end{align}
 and
\begin{align}\label{proof_coop_eq_moment}
\moment\approx P_{T}^n \mathbb{E}_G\left\{G^n\right\}\mathbb{E}_R\left\{F_H\left(\frac{\gamma R^\alpha}{P_{T_S}}\right)\left(\frac{\left|\region-\pi r_c^2\right|}{\left|\region\right|}
 +\frac{\pi r_c^2}{\left|\region\right|}F_H\left(\frac{\gamma R^\alpha}{P_{T_S}}\right)\right)^{M-1}R^{-n\alpha}\right\}.
\end{align}
Hence, we arrive at the results in Theorem~\ref{theorem_coop} and Corollary~\ref{theorem_coop}.
\else
Then the MGF and $n$-th for the interference from a random SU can be obtained by substituting~\eqref{active_coop} into~\eqref{general_mgf} and~\eqref{general_moment} respectively, which can be expressed as~\eqref{proof_coop_eq} and~\eqref{proof_coop_eq_moment} shown at the top of next page.
\begin{figure*}[!t]
\normalsize
 \begin{align}\label{proof_coop_eq}
 \mgf \approx & \int_0^{\infty}\int_{\epsilon}^{r_{\textrm{max}}}\exp\left(-s P_{T} g r^{-\alpha}\right)F_H\left(\frac{\gamma r^\alpha}{P_{T_S}}\right)\left(\frac{\left|\region-\pi r_c^2\right|}{\left|\region\right|}
 +\frac{\pi r_c^2}{\left|\region\right|}F_H\left(\frac{\gamma r^\alpha}{P_{T_S}}\right)\right)^{M-1}\fr\fg dr dg  \nonumber\\
 &+1-\int_{\epsilon}^{r_{\textrm{max}}}F_H\left(\frac{\gamma r^\alpha}{P_{T_S}}\right)\left(\frac{\left|\region-\pi r_c^2\right|}{\left|\region\right|}
 +\frac{\pi r_c^2}{\left|\region\right|}F_H\left(\frac{\gamma r^\alpha}{P_{T_S}}\right)\right)^{M-1}\fr dr.
 \end{align}
\hrulefill
\vspace*{4pt}
\end{figure*}
\begin{figure*}[!t]
\normalsize
\begin{align}\label{proof_coop_eq_moment}
\moment\approx P_{T}^n \mathbb{E}_G\left\{G^n\right\}\mathbb{E}_R\left\{F_H\left(\frac{\gamma R^\alpha}{P_{T_S}}\right)\left(\frac{\left|\region-\pi r_c^2\right|}{\left|\region\right|}
 +\frac{\pi r_c^2}{\left|\region\right|}F_H\left(\frac{\gamma R^\alpha}{P_{T_S}}\right)\right)^{M-1}R^{-n\alpha}\right\}.
\end{align}
\hrulefill
\vspace*{4pt}
\end{figure*}
Hence, we arrive at the results in Theorem~\ref{theorem_coop} and Corollary~\ref{theorem_coop}.
\fi
\end{proof}

\subsection{Proof of Theorem~\ref{theo_meannode}}\label{app_mean_node}
 \begin{proof}
From definition, $\meannode$ is the mean value of the number of active SUs, after averaging over all possible networking realizations. Let $P_{\textrm{active}}$ denote the (unconditional) probability of a SU being active. Mathematically, the average number of active SUs can be written as
  \ifCLASSOPTIONonecolumn
  \begin{align}\label{meannode_define}
\meannode=M\times P_{\textrm{active}}=M\times \mathbb{E}_{R}\{P_{\textrm{active}|R}\}.
\end{align}
\else
  \begin{align}\label{meannode_define}
\meannode=&M\times P_{\textrm{active}}\nonumber\\
=&M\times \mathbb{E}_{R}\{P_{\textrm{active}|R}\}.
\end{align}
\fi

In order to further simplify~\eqref{meannode_define}, we can exploit the fact that $P_T^0=1$, $\mathbb{E}_{G}\{G^0\}=1$ and $R^0=1$. Thus, we can rewrite~\eqref{meannode_define} as
  \begin{align}\label{meannode_define2}
\meannode=&M\times 1\times1\times\mathbb{E}_{R}\{P_{\textrm{active}|R}\times 1\}\nonumber\\
=&M\times\underbrace{ P_T^0 \mathbb{E}_{G}\{G^0\}\mathbb{E}_{R}\{P_{\textrm{active}|R}\times R^0\} }_{\mu_{I}(0)}.
\end{align}

Comparing the latter term in~\eqref{meannode_define2} with~\eqref{general_moment}, we can see that it is in fact the zero-th moment of the interference from a random SU. Thus, we arrive at the result in~\eqref{mean_node}.
\end{proof}

%%****************************************************************************************

%%%%----------------------------------------------------------------------

%%\begin{figure}
%%\centering
%%\includegraphics[width=0.8  \textwidth]{fig/a}
%%        \caption{Comparison with Fig3 in Torriei's paper}
%%        \label{disk_com}
%%\end{figure}
%
\ifCLASSOPTIONpeerreview
\renewcommand{\baselinestretch}{1.78}
\fi

%% \bibliographystyle{IEEEtran}

\end{document}